\documentclass[journal]{IEEEtran}
\ifCLASSINFOpdf \else \fi

\usepackage{graphicx}

\usepackage[cmex10]{amsmath}
\usepackage{subfigure}
\usepackage{caption}
\usepackage{subcaption}
\usepackage{amssymb}
\usepackage{bbm}

\usepackage{xcolor}
\usepackage{tabularx}
\usepackage{acro}[=v3] 
\DeclareAcronym{6G}{short = 6G, long  = sixth generation ,tag = abbrev}
\DeclareAcronym{5G}{short = 5G, long  = 5th generation ,tag = abbrev}
\DeclareAcronym{ISAC}{short = ISAC , long  = integrated sensing and communication,tag = abbrev}
\DeclareAcronym{MMSE}{short = MMSE , long  = minimum mean squared error ,tag = abbrev}
\DeclareAcronym{RMSE}{short = RMSE , long  = root mean squared error ,tag = abbrev}
\DeclareAcronym{DFT}{short = DFT, long  = discrete Fourier transform ,tag = abbrev}
\DeclareAcronym{IDFT}{short = IDFT, long  = inverse discrete Fourier transform ,tag = abbrev}
\DeclareAcronym{FFT}{short = FFT, long  = fast Fourier transform ,tag = abbrev}
\DeclareAcronym{ISFFT}{short = ISFFT, long  = inverse symplectic finite Fourier transform ,tag = abbrev}
\DeclareAcronym{IFFT}{short = IFFT, long  = inverse fast Fourier transform ,tag = abbrev}
\DeclareAcronym{SFFT}{short = SFFT, long  = symplectic finite Fourier transform ,tag = abbrev}
\DeclareAcronym{V2X}{short = V2X, long  = vehicle-to-everything ,tag = abbrev}
\DeclareAcronym{MSE}{short = MSE, long  = mean square error ,tag = abbrev}
\DeclareAcronym{CSI}{short = CSI , long  = channel state information ,tag = abbrev}
\DeclareAcronym{MIMO}{short = MIMO, long  = multiple-input multiple-output ,tag = abbrev}
\DeclareAcronym{BW}{short = Bw, long  = bandwidth ,tag = abbrev}
\DeclareAcronym{LS}{short = LS, long  = least square ,tag = abbrev}
\DeclareAcronym{SNR}{short = SNR, long  = signal-to-noise ratio ,tag = abbrev}
\DeclareAcronym{NMSE}{short = NMSE, long  = normalized mean square error ,tag = abbrev}
\DeclareAcronym{PAPR}{short = PAPR, long  = peak to average power ratio ,tag = abbrev}
\DeclareAcronym{ISI}{short = ISI, long  = inter-symbol interference ,tag = abbrev}
\DeclareAcronym{ICI}{short = ICI, long  = inter-carrier interference ,tag = abbrev}
\DeclareAcronym{CP}{short = CP, long  = cyclic prefix ,tag = abbrev}
\DeclareAcronym{CPP}{short = CPP, long  = chirp-periodic prefix ,tag = abbrev}
\DeclareAcronym{ZP}{short = ZP, long  = zero padding ,tag = abbrev}
\DeclareAcronym{ZS}{short = ZS, long  = zero suffix ,tag = abbrev}
\DeclareAcronym{ZF}{short = ZF, long  = zero forcing ,tag = abbrev}
\DeclareAcronym{RZP}{short = RZP, long  = reduced-zero padded ,tag = abbrev}
\DeclareAcronym{BER}{short = BER, long  = bit error rate ,tag = abbrev}
\DeclareAcronym{SIC}{short = SIC, long  = successive interference cancellation ,tag = abbrev}
\DeclareAcronym{AWGN}{short = AWGN, long  = additive white Gaussian noise,tag = abbrev}
\DeclareAcronym{SINR}{short = SINR, long  = signal-to-interference-plus-noise ratio ,tag = abbrev}
\DeclareAcronym{BPSK}{short = BPSK, long  = binary phase shift keying ,tag = abbrev}
\DeclareAcronym{QPSK}{short = QPSK, long  = quadrature phase shift keying ,tag = abbrev}
\DeclareAcronym{PDF}{short = PDF, long  = probability density function ,tag = abbrev}
\DeclareAcronym{1D}{short = 1D, long  = 1-dimensional ,tag = abbrev}
\DeclareAcronym{2D}{short = 2D, long  = 2-dimensional ,tag = abbrev}
\DeclareAcronym{3D}{short = 3D, long  = three-dimensional ,tag = abbrev}
\DeclareAcronym{QAM}{short = QAM, long  = quadrature amplitude modulation ,tag = abbrev}
\DeclareAcronym{CE}{short = CE, long  = channel estimation ,tag = abbrev}
\DeclareAcronym{FMCW}{short = FMCW, long  = frequency modulated continuous wave ,tag = abbrev}
\DeclareAcronym{JSAC}{short = JSAC, long  = joint sensing and communication ,tag = abbrev}
\DeclareAcronym{OTFS}{short = OTFS, long  = orthogonal time-frequency space ,tag = abbrev}
\DeclareAcronym{SE}{short = SE, long  =spectral efficiency,tag = abbrev}
\DeclareAcronym{MP}{short = MP, long  = message passing ,tag = abbrev}
\DeclareAcronym{OMP}{short = OMP, long  = orthogonal matching pursuit ,tag = abbrev}
\DeclareAcronym{ML}{short = ML, long  = maximum likelihood ,tag = abbrev}
\DeclareAcronym{OFDM}{short = OFDM, long  = orthogonal frequency division multiplexing ,tag = abbrev}
\DeclareAcronym{AFDM}{short = AFDM, long  = affine frequency division multiplexing ,tag = abbrev}
\DeclareAcronym{OCDM}{short = OCDM, long  = orthogonal chirp division multiplexing ,tag = abbrev}
\DeclareAcronym{ODDM}{short = ODDM, long  = orthogonal delay-Doppler division multiplexing,tag = abbrev}
\DeclareAcronym{FD}{short = FD, long  = full duplex,tag = abbrev}
\DeclareAcronym{PRI}{short = PRI, long  = pulse repetition interval,tag = abbrev}
\DeclareAcronym{LTV}{short = LTV, long  = linear time-varying  ,tag = abbrev}
\DeclareAcronym{FrFT}{short = FrFT, long  = fractional Fourier transform  ,tag = abbrev}
\DeclareAcronym{IDAFT}{short = IDAFT, long  = inverse discrete affine Fourier transform,tag = abbrev}
\DeclareAcronym{AFT}{short = AFT, long  = Affine Fourier transform,tag = abbrev}
\DeclareAcronym{STFT}{short = STFT, long  = short-time Fourier transform,tag = abbrev}
\DeclareAcronym{DD}{short = DD-coupling, long  = Doppler-delay coupling,tag = abbrev}
\DeclareAcronym{SBL}{short = SBL, long  = sparse bayesian learning,tag = abbrev}
\DeclareAcronym{EPA}{short = EPA, long  = embedded pilot-aided,tag = abbrev}
\DeclareAcronym{PCTD}{short = PCTD, long  = post-coded time domain,tag = abbrev}
\DeclareAcronym{MRC}{short = MRC, long  = maximum ratio combiner,tag = abbrev}
\DeclareAcronym{RRC}{short = RRC, long  = root raised cosine,tag = abbrev}
\DeclareAcronym{EVM}{short = EVM, long  = error vector magnitude,tag = abbrev}
\DeclareAcronym{EVA}{short = EVA, long  = extended vehicular A,tag = abbrev}
\DeclareAcronym{SI}{short = SI, long  = self interference,tag = abbrev}
\DeclareAcronym{CSC}{short = CSC, long  = circularly shifted chirp,tag = abbrev}
\DeclareAcronym{LFM}{short = LFM, long  = linear frequency modulated,tag = abbrev}
\DeclareAcronym{LoRa}{short = LoRa, long  = long range,tag = abbrev}
\DeclareAcronym{Tx}{short = Tx, long  = transmitting,tag = abbrev}
\DeclareAcronym{Rx}{short = Rx, long  = receiving,tag = abbrev}
\DeclareAcronym{RCS}{short = RCS, long  = radar cross section,tag = abbrev}
\DeclareAcronym{CFO}{short = CFO, long  = carrier frequency offset,tag = abbrev}
\DeclareAcronym{CPI}{short = CPI, long  = coherent processing interval,tag = abbrev}
\DeclareAcronym{WSS}{short = WSS, long  = wide sense stationary,tag = abbrev}
\DeclareAcronym{IID}{short = i.i.d., long  = independent and identically distributed,tag = abbrev}
\DeclareAcronym{NOMA}{short = NOMA, long  = non orthogonal multiple access,tag = abbrev}
\DeclareAcronym{RDM}{short = RDM, long  = range-Doppler map,tag = abbrev}
\DeclareAcronym{LoS}{short = LoS, long  = line of sight,tag = abbrev}
\usepackage{subcaption}
\usepackage{caption}
\usepackage{algpseudocode}
\usepackage[linesnumbered,ruled,vlined]{algorithm2e}
\usepackage{nomencl}
\usepackage{array}
\usepackage{url}
\usepackage{cite}

\begin{document}
\bstctlcite{IEEEexample:BSTcontrol}
\title{Enabling Full Duplex ISAC Leveraging Waveform Domain Separability}

\author{
\IEEEauthorblockN{Abdelali Arous,~\IEEEmembership{Graduate~Student~Member,~IEEE}, Hamza Haif,~\IEEEmembership{Graduate~Student~Member,~IEEE},  and H\"{u}seyin Arslan,}
\IEEEmembership{Fellow, IEEE}
\thanks{The authors are with the Department of Electrical and Electronics Engineering, Istanbul Medipol University, Istanbul, 34810, Turkey (e-mail: abdelali.arous@std.medipol.edu.tr; hamza.haif@std.medipol.edu.tr; huseyinarslan@medipol.edu.tr).}
}


\maketitle

\begin{abstract}
Integrated sensing and communication (ISAC) in monostatic in-band full-duplex (IBFD) systems encounters significant challenges due to self-interference (SI) at the radar receiver during concurrent communication and radar operations. This paper proposes a novel waveform-domain self-interference cancellation (SIC) technique that leverages the unique properties of orthogonal frequency division multiplexing (OFDM) and affine frequency division multiplexing (AFDM) signals. The proposed approach designs the integrated dual-functionality frame to utilize OFDM for communication and AFDM for radar sensing, both generated using the same modulator block. Then, we establish the conditions under which a wide sense stationary (WSS) process in the time domain appears as WSS in the affine domain and demonstrate that the interfering OFDM signal behaves as an additive white Gaussian noise (AWGN) in this domain. Exploiting this property, the received signal is projected into the affine domain, where the SI appears as AWGN, enabling its subtraction with minimal residual interference. To further mitigate the residual SI, an iterative low-complexity windowing scheme is applied, selectively locking onto the radar signal to reduce the processed signal space. A subsequent time domain spreading step is applied after converting the SIC-processed signal into the post-coded time domain, wherein the SI diminishes separately across the delay and Doppler axes. The proposed method demonstrates superior performance in terms of detection probability, target's range and velocity root mean square error (RMSE), while maintaining high spectral efficiency and minimal computational complexity. 
\end{abstract}

\begin{IEEEkeywords}
ISAC, full-duplex, successive interference cancellation (SIC), waveform design, OFDM, AFDM.
\end{IEEEkeywords}

\IEEEpeerreviewmaketitle
\vspace{-1mm}
\section{Introduction}
\IEEEPARstart{A}{t} the advent of each generation of wireless communication networks, spectrum scarcity remains a critical factor shaping the associated technologies. The optimization of spectrum utilization and the effective management of interference are paramount considerations in the design of emerging wireless systems \cite{zhang20196g}. As the standardization of the sixth-generation (6G) network progresses, spectrum sharing has garnered significant attention, particularly within the framework of \ac{ISAC}. \ac{ISAC} aims to unify sensing and communication functionalities within a single network infrastructure, facilitating resource sharing, hardware unification, and energy efficiency \cite{parssinen2021white}. A fundamental aspect of resource sharing in \ac{ISAC} systems is the joint utilization of spectral resources for both radar and communication operations. Despite the migration to higher frequency bands that offer increased bandwidth availability, the exponential growth in connectivity demands and capacity optimization requirements necessitate efficient resource allocation strategies. Furthermore, the coexistence of data transmission and environmental sensing imposes rigorous design constraints on waveform selection as the performance requirements differ substantially \cite{alsaedi2023spectrum}. Consequently, there is an increasing preference toward designing waveforms capable of supporting both functionalities without a major performance compromise. 
\par A feasible approach to achieve \ac{ISAC} is the implementation of \ac{FD} systems, which provide concurrent transmission and reception of sensing and communication signals \cite{barneto2021full, smida2023full}. \ac{FD}-\ac{ISAC} systems have numerous benefits, such as improved spectrum efficiency, reduced latency, and efficient resource allocation. However, a major challenge associated with FD operation is the presence of powerful \ac{SI}, which arises when the communication node transmits data frames while the radar simultaneously receives target echoes. This interference drastically diminishes the quality of the radar signal in monostatic systems. From a standardization perspective, FD operation is being explored within the context of sub-band FD (SBFD), which facilitates concurrent transmission and reception within different sub-bands of the same frequency band. In contrast, in-band FD (IBFD) utilizes the entire band for dual operations, necessitating \ac{SI} cancellation (SIC) to levels below the receiver's noise floor. Consequently, extensive research efforts have been devoted to SIC and mitigation techniques for IBFD systems \cite{hong2014applications, sabharwal2014band, miridakis2012survey, hong2022frequency}. Mainly, SIC techniques in IBFD systems are classified into passive and active approaches. Passive SIC relies on electromagnetic characteristics to separate the transmit and receive chains, thereby attenuating \ac{SI} prior to reception. This includes transmit and receive beamforming, cross-polarization, and physical antenna separation \cite{nwankwo2017survey}. For instance, in \cite{liu2023joint}, a joint design of transmit and receive beamformers was proposed to simultaneously maximize radar beampattern power at the target while suppressing residual SI through a penalty-based iterative optimization algorithm. Similarly, \cite{li2024full} presented a joint transmit and receive beamforming framework for FD-non-orthogonal multiple access (NOMA) ISAC systems, where the optimization aimed to maximize the sensing \ac{SINR} for single and multi-target detection under both perfect and imperfect \ac{CSI}. Moving to mmWave frequencies, \cite{islam2022integrated} proposed an integrated optimization approach for designing analog and digital beamformers in conjunction with SIC within a unified waveform framework. Additionally, cross-polarized antennas can electromagnetically isolate transmit and receive antennas, thereby reducing \ac{SI} \cite{le2021analog}. Nonetheless, although passive SIC techniques provide substantial SI suppression, they also attenuate desired signals and result in a residual SI component. Moreover, for compact monostatic ISAC systems, antenna separation is often infeasible due to spatial constraints especially at high frequencies.
 \par Conversely, active SIC techniques operate by removing a replica of the transmitted signal from the received signal. When this subtraction occurs before analog-to-digital conversion (ADC), it is referred to as analog active SIC. However, since the received signal undergoes distortions due to the nonlinearities of RF components, residual SI persists after analog SIC and must be further suppressed in the digital domain. A frequency-domain SIC approach for \ac{OFDM} systems was proposed in \cite{hong2022frequency}, utilizing near-optimal interference-canceling filter weights to track channel variations and reconstruct the \ac{SI}. Likewise,  \cite{duan2024frequency} introduced a frequency-domain differential interference cancellation (F-DIC) strategy for IBFD-ISAC systems employing \ac{OFDM}, leveraging the invariant characteristics of interference. Following the digital domain SIC, \cite{bernhardt2018self} explores a receiver structure in which the received signal is non-uniformly sampled at the zero-crossing instants of the SI signal to relax the constraints of dynamic range of ADC. Furthermore, \cite{fuchs2024optimized} investigated nonlinear SIC by formulating an optimization problem to minimize estimation uncertainty under practical power limitations. Multi-tap RF cancellers \cite{kolodziej2016multitap}, further exploit the multipath nature of the \ac{SI} channel by aligning RF cancellation taps with the actual channel taps, though this approach introduces additional complexity, particularly when the estimated and actual channel taps are misaligned. Collectively, realistic considerations suggest that no solitary SIC approach; whether passive or active, analog or digital can fully mitigate \ac{SI} in IBFD-ISAC systems. Instead, a hybrid SIC framework integrating multiple cancellation techniques across different domains is necessary for effective \ac{SI} suppression \cite{zhang2022band}.
\par Regarding waveform design for ISAC, the quest for a singular waveform for both sensing and communication in ISAC has garnered considerable attention \cite{ zhang2024intelligent}. Multicarrier waveforms, including OFDM, \ac{OTFS}, and \ac{AFDM}, exhibit adaptability to diverse channel conditions, offering design flexibility and performance improvements in target detection and communication throughput \cite{koivunen2024multicarrier}. Additionally, flexible resource allocation enables the assignment of specific temporal and spatial blocks for radar and communication functions, extending across several domains, including the delay-Doppler and affine domains \cite{rou2024orthogonal}, providing an improved interpretation of channel reflections in those domains. While joint \ac{ISAC} waveform design remains a viable approach \cite{liu2024ofdm, bemani2024integrated}, alternative methods advocate for the transmission of distinct waveforms tailored to each functionality. Specifically, chirp or pulse-like waveforms are often employed for target detection to enhance performance and minimize leakage and interference between sensing pulses and communication signals due to power disparities \cite{xiao2022waveform, csahin2020multi}. However, generating separate waveforms introduces additional complexity, as distinct transmission and reception chains are required for independent signal processing. 
\par In response to these challenges, this paper proposes a novel waveform-domain SIC technique for IBFD-ISAC systems. The proposed scheme leverages the generalized modulation and demodulation framework of \ac{AFDM} to generate a chirp-like signal for sensing and an OFDM-based frame for communication \cite{bemani2023affine}. By exploiting the flexibility of the \ac{AFDM} structure, different waveform configurations can be generated through post- and pre-processing phases of the \ac{OFDM} demodulator. Specifically, after analog SIC, the residual \ac{SI}-affected \ac{OFDM} signal is projected onto the affine domain, where it exhibits a noise-like spreading behavior, whereas the \ac{AFDM} radar signal remains more localized. This property is exploited for \ac{SI} suppression by applying an iterative windowing algorithm to mitigate residual \ac{SI}, yielding an enhanced radar signal. Furthermore, an additional spreading process is applied in the time domain after converting the SIC-processed \ac{AFDM} signal to the \ac{PCTD}, thereby reducing phase variations along the delay and Doppler axes \cite{arousnovel}. This two-stage waveform-domain SIC approach provides an additional separation mechanism that facilitates signal purification within their respective multiplexing domains, significantly improving \ac{SI} suppression while maintaining a compact signal generation and processing chain. To the best of our knowledge, this is the first attempt to perform SIC using the projections in waveform-domains. The main contributions of this paper are summarized as follows:
\begin{itemize}
        \item Based on the generalized generation framework of multicarrier waveforms, we proposed an IBFD-\ac{ISAC} system that interchangeably utilizes \ac{OFDM} and \ac{AFDM} waveforms for communication and sensing, respectively, by tuning different design parameters and resource elements.
        \item We analytically characterize the statistical behavior of any \ac{WSS} time domain signals in the affine domain and demonstrate that under specific conditions, \ac{OFDM} exhibits \ac{AWGN} properties in this domain. Hence, at certain segments of the received radar signal, the \ac{SI} and the thermal receiver noise can be jointly approximated as \ac{AWGN} with constant power levels. 
        \item Capitalizing on this statistical property, the \ac{SI} originated from \ac{OFDM} signal can be reconstructed and subtracted in the affine domain where its has a predictive behavior, followed by an iterative windowing scheme for residual SI suppression. The proposed windowing algorithm drastically reduces the processed frame size where it locks on the least and maximum indexed target bins. Then the SIC-processed signal is taken to the time domain to perform an additional spreading phase, where any phase or amplitude variation is spreaded in \ac{PCTD} before performing target detection. 
        \item We provide a comprehensive performance analysis of the proposed IBFD-ISAC system, evaluating its \ac{SINR}, probability of detection, computational complexity, and spectral efficiency under varying radar and communication configurations.  
\end{itemize}
The rest of this paper is organized as follows: Section II presents the waveform design for the monostatic IBFD-ISAC system along with the received signal model. Section III discusses the affine-domain representation of \ac{OFDM} signals. Following that, the proposed waveform-domain SIC is presented along with the iterative windowing algorithm and the performance analysis in Section IV. Simulation results and the conclusion are given in Section V and VI, respectively.            
\par \textit{Notation}: Bold uppercase $\mathbf{A}$, bold lowercase $\mathbf{a}$, and unbold letters $A,a$ denote matrices, column vectors, and scalar values, respectively. $\text{tr}(\mathbf{A})$, $\mathbbm{1}(\cdot)$ and $(\cdot)^{-1}$ denote the trace of $\mathbf{A}$, the indicator function and the inverse operator, respectively. $\delta(\cdot)$, $\mathbb{E}(\cdot)$ and $\Pi\left(\frac{t}{T}\right)$ denote the Dirac-delta function, the expectation and rectangular function of duration $T$. $\operatorname{diag}\left( {A}_1,\dots,{A}_{N} \right)$ returns the diagonal matrix composed of ${A}_1,\dots,{A}_{N}$ on its diagonal, respectively. $\mathbb{C}^{{M\times N}}$ denotes the space of $M\times N$ complex-valued matrices. $\mathbf{A} \odot \mathbf{B}$ is the Hadamard product of $\mathbf{A}$ and $\mathbf{B}$ and symbol $j$ represents the imaginary unit of complex numbers with $j^2=-1$.

\section{The Proposed System Model}
In this section, the system model of the proposed monostatic IB\ac{FD}-\ac{ISAC} transceiver is detailed. The objective is to design and transmit orthogonal and distinct waveforms for sensing and communication while mitigating the \ac{SI} superimposed on the targets echos at the radar receiver, as illustrated in Fig. \ref{fig: block_diagram}. Unlike the continuous wave radar, the proposed transmission scheme follows the pulsed radar operation mode where the short duration sensing signal is transmitted periodically in each \ac{PRI} while the communication signal is transmitted in the remaining duration coinciding with the silent period of the radar. \textcolor{black}{This separate design of sensing and communication waveforms, while reducing the achievable \ac{SE}, can transmit different waveforms using the same generation block for the following reasons: 
\begin{itemize}
    \item Random communication signals has less focused ambiguity function leading to reduced matched filtering performance, and lower detection probability at low SNR. Moreover, high-order modulation and number of subcarriers create non-constant envelopes with high PAPR, limiting the amplification efficiency.
    \item Alternatively, the design AFDM sensing signal benefits from deterministic and constant-envelope structure which allows efficient power amplification and sharper detection. Unlike communication signals, it is optimized for power allocation, modulation, and spectral shaping.
    \item A guard interval is appended to reduce the inter-frame leakage due the phase discontinuity between AFDM and OFDM signals. Note that this guard is very small and is not required for the interference rejection.
    \item This separated design is then used to perform SIC using the different waveform domain properties, rather than the classical power-domain cancellation. 
\end{itemize}
}
The following subsection describes the transmission frames. 
\vspace{-2mm}
\subsection{Frame Design}
\ac{AFDM} is a generalization of the chirp-based multicarrier waveforms, where the well-known \ac{FMCW} radar signal using linear chirp signals falls into. However, rather than transmitting a single unmodulated chirps, \ac{AFDM} multiplexes multiple chirps in the affine domain, those \ac{CSC} chirps are used to carry modulated data and can be converted to time-domain using  \ac{IDAFT}. Given a data vector $x_r[m]$ modulated within the sensing signal in the affine domain with $m = 0, \cdots, N_r-1$ and $N_r$ as the total number of subcarriers, the \ac{AFDM} signal spanning a bandwidth $B$ can be written as 
\begin{equation}
    \begin{aligned}
    s^r[n] = \frac{1}{\sqrt{N_r}} \sum_{m=0}^{N_r-1} x_r[m] e^{j2\pi(c_1n^2+\frac{mn}{N_r}+c_2m^2)},   
        \label{equ: AFDM_signal}
    \end{aligned}
\end{equation}
where $c_1$ and $c_2$ are the refining coefficients of \ac{AFDM} chosen to accommodate the channel variations \cite{bemani2023affine}. In matrix form, \eqref{equ: AFDM_signal} can be written as $\mathbf{s}^r = \boldsymbol{\Lambda}^{H}\mathbf{x}_r $ where $\boldsymbol{\Lambda}$ is the DAFT unitary matrix, and can be decomposed into a DFT matrix $\boldsymbol{F}$ with two diagonal quadratic matrices $\boldsymbol{\Lambda}_{c_1}$ and $\boldsymbol{\Lambda}_{c_2}$ such as $\boldsymbol{\Lambda} = \boldsymbol{\Lambda}_{c_1} \mathbf{F} \boldsymbol{\Lambda}_{c_2}$, where $\boldsymbol{\Lambda}_{c_{\{1,2\}}} = \text{diag}(e^{-j2\pi c_{\{1,2\}}n^2}, n = 0,1,\dots,N_r-1)$. For a total bandwidth $B$, the symbol duration for one \ac{AFDM} symbol is defined as $T_r = \frac{N_r}{B}$ and is assumed to be smaller than the round trip delay of the radar signals. The $N_r$ carriers contain $N_r^p$ pilot subcarrier used for radar processing which can be adjusted as will be shown in Section V-D. The remaining $N_r-N_r^p$ carriers can be used for conventional data transmission for systems using \ac{AFDM} modulation. To conserve the periodicity of \ac{AFDM}, a specific condition must be met after appending the \ac{CPP} such as 
\begin{equation}
    \begin{aligned}
        s^r[n] = s^r[N_r+n] e^{-j2\pi c_1 (N_r^2+2N_rn)}, n = -L_{cpp},\dots, -1,
        \label{equ: phase_AFDM}
    \end{aligned}
\end{equation}
where $L_{cpp}$ is the size of the \ac{CPP} (of duration $T_{cpp}$). Note that when $N$ is even and $2Nc_1$ is integer, \ac{CPP} is equivalent to the conventional \ac{CP}.
\par
\begin{figure}[t!]
    \centering  \includegraphics[width=0.5\textwidth]{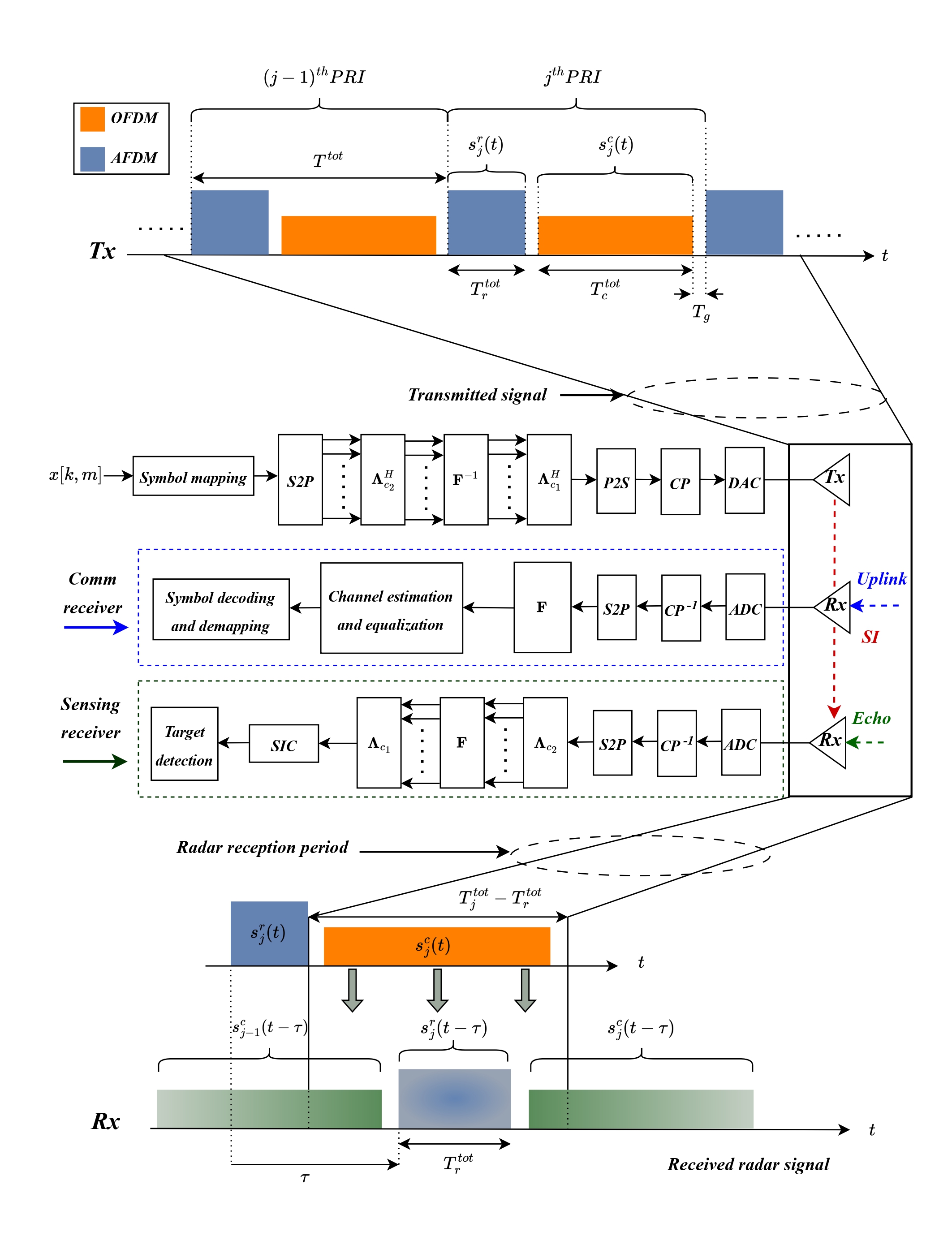}
    \caption{\textcolor{black}{Block diagram of the proposed frame generation based on OFDM modulator.}}
    \label{fig: block_diagram}
\end{figure}

 Alternatively, for communication users, the conventional \ac{OFDM} signal with $N_c$ data subcarriers having a symbol duration of $T_c = \frac{N_c}{B}$ is used, given by
\begin{equation}
    \begin{aligned}
     s^c[n] = \frac{1}{\sqrt{N_c}}\sum_{k=0}^{N_c-1} x_c[k] e^{j2\pi\frac{kn}{N_c}}, 
        \label{equ: OFDM_signal}
    \end{aligned}
\end{equation}
with $x_c[k]$ representing the data vector and $k = 0, \cdots, N_c$ indicating the indices in the frequency domain. After appending a \ac{CP} duration $T_{cp}$ of size $ L_{cp}\geq \ L_{max} $, where $L_{max}$ is the maximum access delay of the communication channel, the \ac{OFDM} signal is given as
\begin{equation}
    \begin{aligned}
     s^c[n]= \begin{cases}s^c\left[n+N_c-L_{\mathrm{cp}}\right], & -L_{\mathrm{cp}} \leq n<0 \\ s^c[n], & 0 \leq n<N_c\end{cases}. 
        \label{equ: OFDM_signal_cp}
    \end{aligned}
\end{equation}
 It can be observed from \eqref{equ: AFDM_signal} and \eqref{equ: OFDM_signal} that by setting $c_1 = c_2 = 0$ \ac{OFDM} can be generated from the \ac{AFDM} signal, in other words, by adjusting different values of the refining coefficients different waveform options can be generated using the same unified block, as shown in Fig. \ref{fig: block_diagram}. Additionally, by activating a single subcarrier in the affine domain, a \ac{LFM} or \ac{LoRa} signal can be transmitted based on the activated carrier index such as
\begin{equation}
    \begin{aligned}
     \mathbf{s}^r_{chirp} = \begin{cases} \boldsymbol{\Lambda}^{H}x_r[0, \cdots, m_0, \cdots, 0]^T , \textit{ LoRa}\\
    \boldsymbol{\Lambda}^{H}x_r[1, 0, \cdots \cdots, 0]^T, 
     \textit{ LFM}    
     \end{cases}, 
        \label{equ: OFDM_signal_1}
    \end{aligned}
\end{equation}
with $m_0$ is the activated chirp carrier index in the affine domain. 
However, due to the interchangeable setting of the refining parameters, the phase continuity between the consecutive \ac{AFDM} and \ac{OFDM} signals is not always guaranteed as highlighted in \eqref{equ: OFDM_signal_cp} and \eqref{equ: phase_AFDM}. Thus, a guard period of duration $T_g =\frac{N_g}{B}$ is left between each radar and communication signals, where $N_g$ is the guard size. Consequently, the transmitted signal at the $j$-th \ac{PRI} in continuous time domain is defined as
\begin{equation}
    \begin{aligned}
        s_j(t) = \begin{cases} s_{j}^r(t),  \quad \quad 0\leq t \leq T_r^{tot}
            \\ 0, \text{ } \text{ }\quad \quad  \quad  T_r^{tot} < t\leq T_r^{tot} +T_g 
            \\  s_{j}^c(t), \quad \quad  T_r^{tot} +T_g  < t \leq T^{tot}
        \end{cases},
        \label{equ: PRI_signal}
    \end{aligned}
\end{equation}
with, 
\begin{equation}
    \begin{aligned}
         \begin{cases} s_{j}^r(t) =  \sum_{\eta=0}^{M_r-1} s_{j}^{r,\eta}(t) \Pi\left(\frac{t- \eta T_r^{\eta}}{T_r^{\eta}}\right)\\  s_{j}^c(t)= \sum_{\eta=0}^{M_c-1} s_{j}^{c,\eta}(t) \Pi\left(\frac{t-\eta T_c^{\eta}}{T_c^{\eta}}\right)
        \end{cases},
    \end{aligned}
\end{equation}
where $T_r^{\eta} = T_r+T_{cpp}$ and $T_c^{\eta} = T_c+T_{cp}$ denoting the total duration of each \ac{AFDM} and \ac{OFDM} symbol, respectively. $T_r^{tot} = M_r T_r^{\eta} $, $T_c^{tot} = M_c T_c^{\eta}$, $M_r$ and $M_c$ are the total duration and the number of symbols of the radar and communication signals at the $j$-th \ac{PRI}, respectively. Then, the total duration of the $j$-th \ac{PRI} is given by
\begin{equation}
    \begin{aligned}
        T^{tot} = T_r^{tot}+T_c^{tot}+2T_g. 
    \end{aligned}
\end{equation}
In matrix form, for a total of $J$ \ac{PRI}s, the transmitted signal is given by $\bold{S}= [\bold{s}_1, \cdots, \bold{s}_j, \cdots, \bold{s}_J]$, where $\bold{s}_j=[\bold{s}_{j}^r \text{ } \bold{g} \text{ } \bold{s}_{j}^c \text{ } \bold{g}]$ and
\begin{equation}
   \begin{cases}
      \bold{s}_{j}^r = [\bold{s}_{j}^{r,0},  \cdots, \bold{s}_{j}^{r,M_r-1}]^T \in \mathbb{C}^{N_r^{tot} \times 1}\\
      \bold{g}= [0, \cdots, 0]^T  \\
      \bold{s}_{j}^c = [\bold{s}_{j}^{c,0},  \cdots, \bold{s}_{j}^{c,M_c-1}]^T \in \mathbb{C}^{ N_c^{tot} \times 1}
   \end{cases},
\end{equation}
with $N_r^{tot} = (N_r+L_{cpp})M_r$ and $N_c^{tot} = (N_c+L_{cp})M_c$. 
\subsection{Received Signal Model}
Before providing the received signal components, the following assumptions about the proposed IB\ac{FD}-\ac{ISAC} system are given
\begin{itemize}
    \item The space domain isolation between the \ac{Tx} and \ac{Rx} antennas is assumed to be within an acceptable level of the ADC's dynamic range, and before passing by the low-power amplifier in order to avoid saturation.
    \item The objects of interests for radar processing are considered to be moving targets with a measurable Doppler shift, whereas the communication reflected signal has a negligible Doppler shift. 
    \item By setting an appropriate cancellation range of the transversal filter, close-range echoes' delays are effectively eliminated. However, \ac{SI} and multipath interference cannot be suppressed due to power difference and the maximum cancellation range constraints, respectively.  
    \item The \ac{Tx} and \ac{Rx} chains are assumed to be perfectly synchronized, and the \ac{CFO} is presumed to be compensated at the radar receiving end. 
    \item \textcolor{black}{The maximum round-trip delay is assumed to be shorter than the \ac{PRI}, such that the target echoes arrive strictly before the next PRI, and thus avoiding range ambiguity.}
\end{itemize}

To this extent, the received signal at the radar silent period of the $j$-th \ac{PRI} coming from $L$ targets can be written as
\begin{equation}
    \begin{aligned}
    r_j(t) =& \sum_{i=1}^L \alpha_i \hat{s}_j(t-\tau_i) e^{j2\pi \nu_i(t-\tau_i)} +\beta s_{j}^c(t)+w_j(t)\\
    & \textit{ for,} \quad  0 \leq t \leq T^{tot}-T_r^{tot},
        \label{equ: recieved_signal}
    \end{aligned}
\end{equation}
with $w_j(t) \sim \mathcal{N}\left(0, \sigma^2\right)$ and $\beta$ denoting the \ac{AWGN} with a variance $\sigma^2$ and the complex channel coefficient of the \ac{SI} link. \textcolor{black}{The complex amplitude of the radar return, delay and Doppler shift related
to the $i$-th path are denoted by $\alpha_i$, $\tau_i$ and $\nu_i$, respectively.} The round-trip delay and Doppler shifts are given as
\textcolor{black}{\begin{equation} \small
     l_i \approx \frac{2R_i}{c} = B\tau_i, \quad \quad \kappa_{i} \approx \frac{2V_if_c}{c} = T_{r} \nu_i,
    \label{equ:range}
\end{equation}}
where $c$ and $f_c$ are the speed of light and the carrier frequency, and  $l_i, \kappa_{i} \in \mathbb{Q}$ denote the sampled delay and Doppler shifts corresponding to the $i$-th target, respectively. For a single target, the signal $\hat{s}_j(t-\tau)$ in \eqref{equ: recieved_signal} is composed of the communication signal from the previous \ac{PRI} $s_{j-1}^c(t)$ and the reflected radar and communication signals of the current \ac{PRI} $s^{r}_j(t)$ and $s_{j}^c(t)$, respectively, as seen in Fig. \ref{fig: block_diagram}. Assuming the communication signal undergoes the same delay shift between two consecutive \ac{PRI}s, the received signal at the radar receiver at the $j$-th \ac{PRI} can be written as
\begin{equation}
    \begin{aligned}
    \hat{s}_j = \begin{cases}
    s_{j-1}^c(t+T_c^{tot}+T_g-\tau), \quad 0\leq t\leq \tau-T_g \\
    s_{j}^r(t-\tau), \quad \quad \quad \quad \quad \quad \tau < t\leq \tau+T_r^{tot} \\
    s_{j}^c(t-\tau-T_r^{tot}-T_g),  \quad  \tau+T_r^{tot}+T_g < t\leq T_j^{tot}\\
    0, \quad \quad \left(\tau-T_g, \tau\right] \cup\left(\tau+T_r^{t o t}, \tau+T_r^{t o t}+T_g\right].
    \end{cases}
        \label{equ: recieved_signal3}
    \end{aligned}
\end{equation}
Hence the received signal can be defined piecewise for $\hat{\alpha} = \alpha e^{-j2\pi \nu \tau}$, as
\begin{equation}
    \begin{aligned}
    r_j(t) = \begin{cases}
    \hat{\alpha} s_{j-1}^c(t+T_c^{tot}+T_g-\tau) e^{j2\pi \nu t} +w_j(t), \\ \quad \textit{ for} \quad \quad   0\leq t\leq T_g \\
    \hat{\alpha} s_{j-1}^c(t+T_c^{tot}+T_g-\tau) e^{j2\pi \nu t} +\beta s^c_j(t)+w_j(t), \\ \quad \textit{ for} \quad \quad   T_g < t\leq \tau-T_g \\
    \beta s^c_j(t)+w_j(t), \textit{ for} \quad \quad   \tau-T_g < t\leq \tau \\
    \hat{\alpha} s_{j}^r(t-\tau) e^{j2\pi \nu t} +\beta s^c_j(t)+w_j(t) , \\  \quad \textit{ for} \quad   \quad \tau < t\leq \tau+T_r^{tot} \\
    \beta s^c_j(t)+w_j(t), \textit{ for} \quad \tau+T_r^{tot} < t\leq \tau+T_r^{tot}+T_g \\
    \hat{\alpha} s_{j}^c(t-\tau-T_r^{tot}-T_g) e^{j2\pi \nu t} +\beta s^c_j(t)+w_j(t),  \\ \quad \textit{ for}    \quad \quad  \tau+T_r^{tot}+T_g < t\leq T^{tot}_j-T_r^{tot}-T_g\\
     \hat{\alpha} s_{j}^c(t-\tau-T_r^{tot}-T_g) e^{j2\pi \nu t} +w_j(t),  \\ \quad \textit{ for}    \quad \quad  T^{tot}_j-T_r^{tot}-T_g < t\leq T^{tot}_j-T_r^{tot}.
    \end{cases}
        \label{equ: recieved_signal4}
    \end{aligned}
\end{equation}
As indicated in \cite{xiao2022waveform}, the Doppler phase term $e^{-j2\pi \nu t}$ can be approximated as $e^{-j2\pi \nu t}\approx 1$ over a short duration \ac{PRI}. This implies that Doppler variations remain undetectable, even for larger values of $M_r$, when $\nu T_r \ll 1$, thereby constraining Doppler information extraction to the total number of $J$ PRIs. However, for \ac{AFDM}, even a small Doppler shift manifests as a scaled shift in the affine domain, given by
\begin{equation}
    \begin{aligned}
    \delta(t-\tau)e^{j2\pi\nu(t-\tau)} \overset{\text{affine}}{\underset{\text{time}}{\rightleftharpoons}} \delta(m-loc)e^{j2\pi \phi(c_{\{1,2\}},l)},
        \label{equ:chirp_channel}
    \end{aligned}
\end{equation}
where $loc = 2Nc_1l+\kappa$ and $\phi(c_{\{1,2\}},l)$ are the equivalent coupled tap shift and the total phase in the affine domain as a function of the delay and refining coefficients, respectively. Therefore, for the proposed \ac{AFDM}-based sensing signal, both the target's delay and Doppler information can be extracted from the symbols within each \ac{PRI} without the need to wait for the whole $J$ \ac{PRI}s, as will be shown in Section IV-B. Therefore, to ensure mathematical tractability and accuracy in the following derivations, the Doppler phase term will be explicitly preserved.
\par After down-converting the signal, the sampled received signal can be written in matrix form as 
\begin{equation}
    \begin{aligned}
    \bold{r}_j = \sum_{i=1}^L \underbrace{\hat{\alpha}_i \bold{\Pi}^{l_i} \bold{\Delta}^{\kappa_i} \hat{\bold{s}}_{j,n_{l_i}}}_{echo} + \underbrace{\beta \bold{s}^c_j}_{ SI} + \bold{w}_j,
        \label{equ: recieved_signal_matrix}
    \end{aligned}
\end{equation}
where $\bold{\Pi}^{l_i}$ is the forward cyclic-shift matrix with entries $\Pi^{l_i}(l,\kappa) = \delta([l-\kappa]_{N}-l_i)$, and $\bold{\Delta}^{\kappa_i}$ is the square diagonal matrix 
\begin{equation}
    \begin{aligned}
    \boldsymbol{\Delta}^{\kappa_i}=\operatorname{diag}\left(e^{-j \frac{2 \pi \kappa_i n}{N}}, n=0,1, \ldots, N-1\right),
        \label{equ: diagonal_matrix}
    \end{aligned}
\end{equation}
Also, for the normalized delay samples $n_{l_i} = B \tau_i$, the vector $\hat{\bold{s}}_{j,n_{l_i}}$ is defined as
\begin{equation}
    \begin{aligned}
       \hat{\bold{s}}_{j,n_{l_i}} =  
       \bigr[& \bold{s}_{j-1}^c[N^{tot}_c+N_g-n_{l_i}], \cdots, \bold{s}_{j-1}^c[N^{tot}_c-1],\bold{g},\bold{s}_j^r, \bold{g},\\ & \bold{s}_j^c[0],\cdots,  \bold{s}_{j}^c[N^{tot}_r-N_g-n_{l_i}-1 ] \bigr]^T,
        \label{equ: recieved_signal_matrix1}
    \end{aligned}
\end{equation}
where $N^{tot} = N_r^{tot}+N_c^{tot}+2N_g $.
\section{Affine-Domain OFDM's Representations}
This section analyzes the statistical properties of a time domain \ac{WSS} signal in the affine domain. Subsequently, the representation of an \ac{OFDM} signal in the affine domain is evaluated for specific conditions. The following analysis is a generalization of the work in \cite{de2024wide} which described the behavior of a discrete-time domain \ac{WSS} signal in the discrete-Fresnel domain.  
\vspace{-6mm}
\subsection{WSS Signal in Affine-Domain}
\begin{figure*}[t]
    \centering
    \subfigure[]{\includegraphics[width=0.32\textwidth]{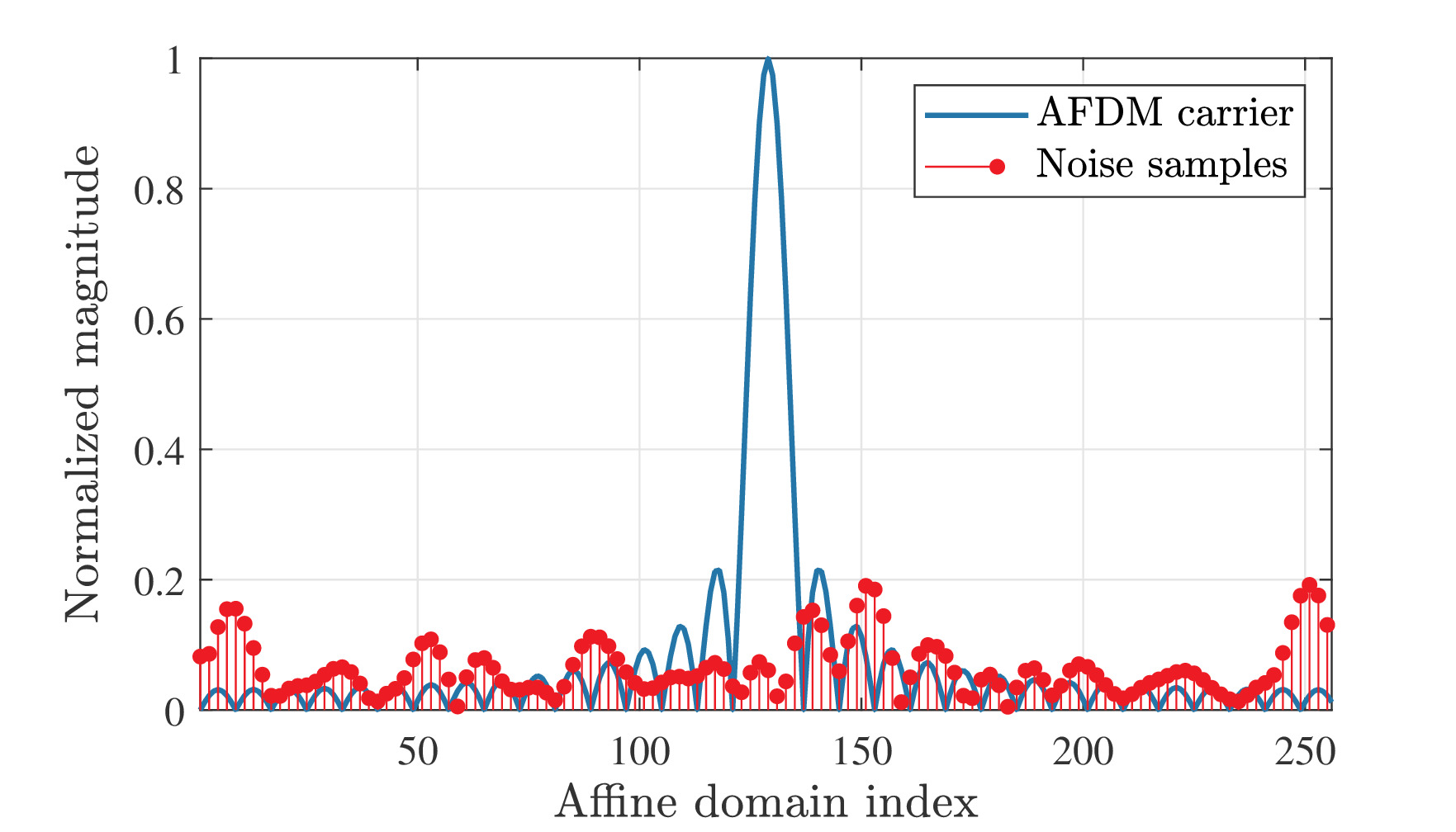}} 
    \hfill
    \subfigure[]{\includegraphics[width=0.32\textwidth]{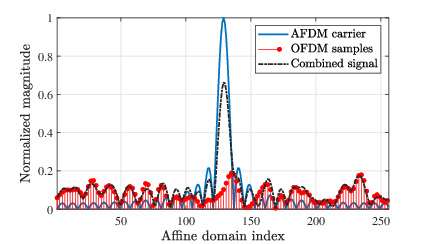}} 
    \hfill
    \subfigure[]{\includegraphics[width=0.32\textwidth]{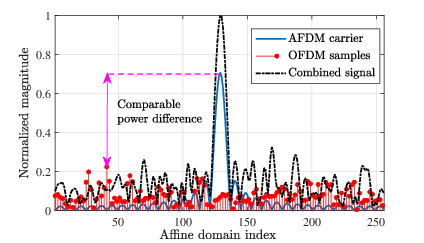}} 
    \hfill
    \subfigure[]{\includegraphics[width=0.32\textwidth]{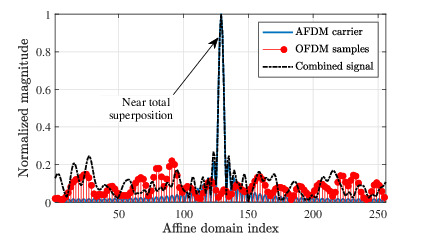}} 
    \hfill
    \subfigure[]{\includegraphics[width=0.32\textwidth]{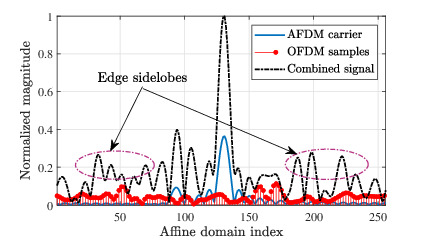}}
    \hfill
    \subfigure[]{\includegraphics[width=0.32\textwidth]{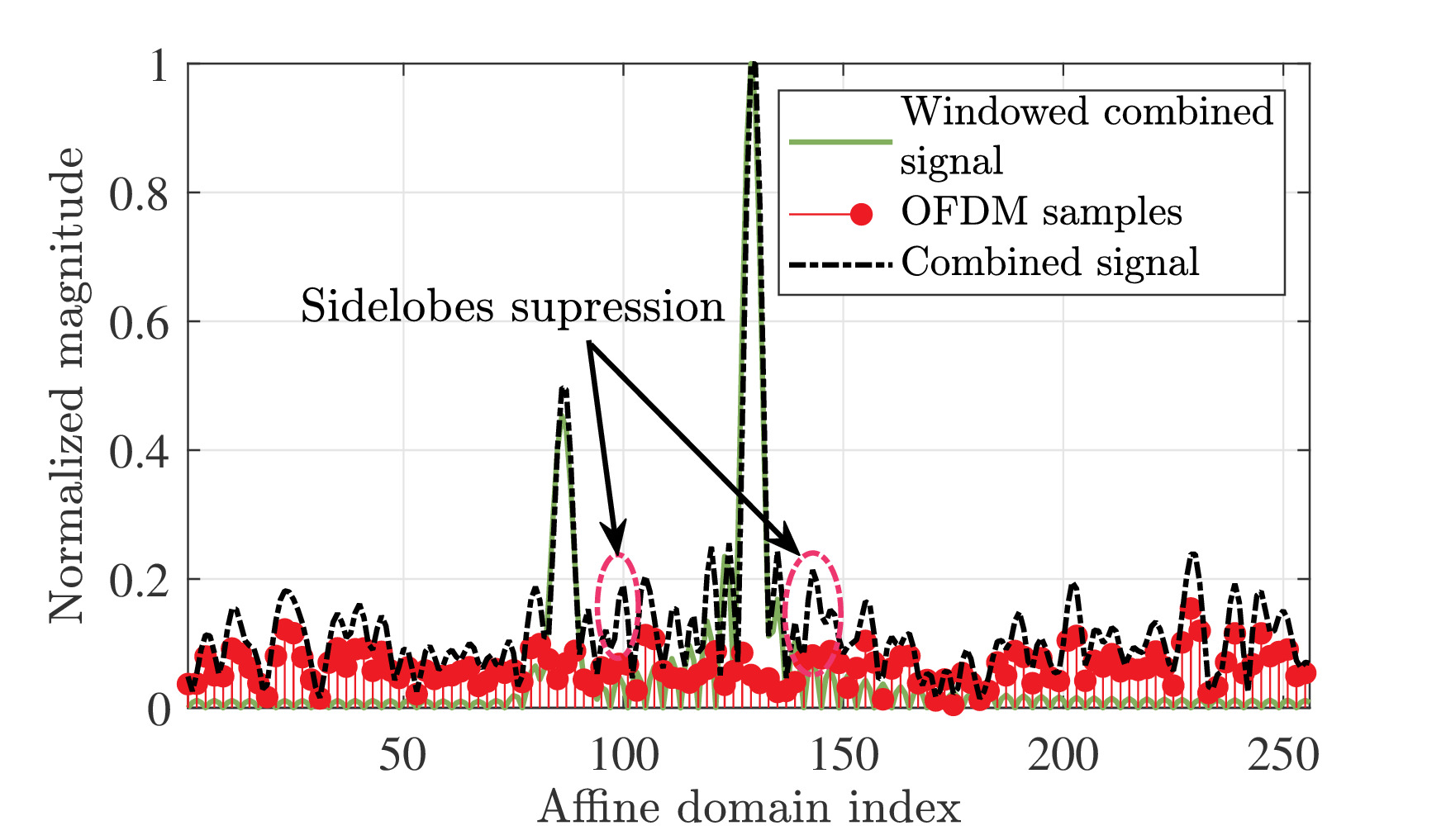}} 
    \caption{Affine-domain representation of: (a) AFDM with AWGN for $\sigma^2 = 0.1$, (b) AFDM with OFDM signal for $N_r = N_c$, (c) AFDM with OFDM signal for $N_r < N_c$, (d) AFDM with OFDM signal for $N_r > N_c$, (e) AFDM with OFDM signal after passing by a two-tap channel, (f) AFDM with OFDM signal after iterative windowing.}
    \label{fig: Affine_domain_representation}
\end{figure*}
Consider a time domain \ac{WSS} signal $r[n]$ defined over discrete indices $n=0,\cdots, N-1$, characterized by a constant expected value given by $\mathbb{E}\left\{r[n]\right\} = \mu_r^t$ and an autocorrelation function expressed as 
$\mathbb{E}\left\{r[n]r^*[\dot{n}]\right\} = R_{rr}^t[n_{\delta}]$, where $n_{\delta}=n-\dot{n}$ denotes the lag between two discrete time indices $n$ and $\dot{n}$, respectively. To observe the statistical behavior of this signal in the affine domain, the expectation and autocorrelation function are calculated for the signal vector $\mathbf{r}$ after applying an $N$-points DAFT such as $\boldsymbol{\Lambda}\mathbf{r}$. Then, the expectation function of a random process in the affine domain is given by 
\begin{equation}
    \begin{aligned}
    \mu_r^a &= \mathbb{E}\left\{\frac{1}{\sqrt{N}}\sum_{n=0}^{N-1}r[n]e^{-j2\pi(c_1n^2+\frac{nm}{N}+c_2m^2)}\right\}\\
    & = \frac{1}{\sqrt{N}}\sum_{n=0}^{N-1}\mathbb{E}\left\{r[n]\right\}e^{-j2\pi(c_1n^2+\frac{nm}{N}+c_2m^2)}\\
    &= \frac{1}{\sqrt{N}}\mu_r^t e^{-j2\pi c_2m^2}\sum_{n=0}^{N-1}e^{-j2\pi (c_1n^2+\frac{nm}{N})}.
        \label{equ:mean_afdm}
    \end{aligned}
\end{equation}
In addition to the phase perturbation term $c_2 m^2$, the expectation function varies across the affine-domain. When selecting $c_1 \ll 1$ with $c_1\in \mathbb{R}^+$, the quadratic phase term becomes negligible, i.e., $c_1 n^2 \approx 0$. Under this condition, the summation can be approximated by a DFT kernel, given by  
$\sum_{n=0}^{N-1}e^{-j2\pi \frac{nm}{N}} = N\delta(m)$. Thus, the expression can be rewritten as
\begin{equation} 
    \begin{aligned}
    \mu_r^a = \sqrt{N}\mu_r^t e^{-j2\pi c_2m^2} \delta(m) = \begin{cases}
        \sqrt{N}\mu_r^t, \quad m = 0\\
        0, \quad \quad \quad m \neq 0
    \end{cases}.
        \label{equ:mean_afdm1}
    \end{aligned}
\end{equation}
Since \( \mu_r^a \) is defined piecewise with respect to \( m \), it follows that the expectation function of a WSS signal is not uniform across the affine domain unless the mean of the time-domain signal satisfies \( \mu_r^t = 0 \). Consequently, for a zero-mean time-domain signal, the corresponding mean in the affine domain remains constant and equals zero. Similarly, the autocorrelation can be calculated as
\begin{equation}
    \begin{aligned}
    & R_{rr}^a = \mathbb{E}\Bigg\{\Bigg(\frac{1}{\sqrt{N}} \sum_{n=0}^{N-1} r[n] 
    e^{-j2\pi\left(c_1n^2 + \frac{nm}{N} + c_2m^2\right)}\Bigg) \notag \\
    & \times \Bigg(\frac{1}{\sqrt{N}} \sum_{\dot{n}=0}^{N-1} r^*[\dot{n}] 
    e^{j2\pi\left(c_1\dot{n}^2 + \frac{\dot{n}\dot{m}}{N} + c_2\dot{m}^2\right)}\Bigg)\Bigg\}\\
    & \quad \quad = \frac{1}{N} \sum_{n=0}^{N-1} \sum_{\dot{n}=0}^{N-1}e^{-j2\pi \Big[c_1(n^2-\dot{n}^2)+\frac{(nm-\dot{n}\dot{m})}{N}+c_2(m^2-\dot{m}^2)\Big]}\\
    &  \times \mathbb{E}\left\{r[n] r^*[\dot{n}]\right\}, 
    \label{equ: auto_afdm}
    \end{aligned}
     \tag{20}
\end{equation}
for $\mathbb{E}\left\{r[n]r^*[\dot{n}]\right\} = R_{rr}^t[n-\dot{n}]$, we \textcolor{black}{can} write
\begin{equation} \setcounter{equation}{21}
    \begin{aligned}
R_{rr}^a &= \frac{1}{N} \sum_{n=0}^{N-1} \sum_{\dot{n}=0}^{N-1} e^{-j2\pi \Big[c_1(n^2-\dot{n}^2)+\frac{(nm-\dot{n}\dot{m})}{N}+c_2(m^2-\dot{m}^2)\Big]}\\
&  \times R_{rr}^t[n-\dot{n}]\\
& = \frac{1}{N} \sum_{\dot{n}=0}^{N-1} e^{j2\pi(c_1\dot{n}^2+\frac{\dot{n}\dot{m}}{N}+c_2\dot{m}^2)} \sum_{n=0}^{N-1} R_{rr}^t[n-\dot{n}] \\
& \times e^{-j2\pi(c_1n^2+\frac{nm}{N}+c_2m^2)}. 
    \label{equ: auto_afdm3}
    \end{aligned}
\end{equation}
It can be seen that $\sum_{n=0}^{N-1} R_{rr}^t[n-\dot{n}] e^{-j2\pi(c_1n^2+\frac{nm}{N}+c_2m^2)}$ is the affine transformation of $R_{rr}^t[n-\dot{n}]$. According to \eqref{equ:chirp_channel}, a shift in time is equivalent to a scaled shift in the affine domain such that $\Lambda R_{rr}^t[n-\dot{n}] \overset{\text{affine}}{\underset{\text{time}}{\rightleftharpoons}} R_{rr}^a[m-2Nc_1\dot{n}]$. Therefore, \eqref{equ: auto_afdm3} can be be written as
\begin{equation}
    \begin{aligned}
R_{rr}^a &=  \frac{1}{\sqrt{N}} \sum_{\dot{n}=0}^{N-1} R_{rr}^a[m-2Nc_1\dot{n}] e^{j2\pi(c_1\dot{n}^2+\frac{\dot{n}\dot{m}}{N}+c_2\dot{m}^2)}. 
    \label{equ: auto_afdm4}
    \end{aligned}
\end{equation}
Before applying the IDAFT, it is important to observe that \( R_{rr}^t[m - 2N c_1 \dot{n}] \) undergoes both scaling and reversal with respect to the summation variable \( \dot{n} \). This necessitates an examination of the transformation properties of the matrix \( \mathbf{\Lambda} \). Notably, since \( \mathbf{\Lambda} \) is composed of two diagonal circulant matrices, the reversal of time-domain samples results in a corresponding reversal in the affine domain, and vice versa \cite{de2024wide}. Mathematically this circular shift property can be described as:
\begin{equation}
    \begin{aligned}
    r[m] = \sum_{n=0}^{N-1} \Lambda (n,m) s[n] \Longleftrightarrow r[-m] = \sum_{n=0}^{N-1}\Lambda (n,m)s[-n].
        \label{equ: reversal}
    \end{aligned}
\end{equation}
Second, although the DAFT is a subset of the linear canonical transforms (LCTs) \cite{pei2000closed}, its scaling invariance property does not hold for arbitrary scaling factors. However, since $c_1$ is typically chosen as $c_1 = \frac{k}{2N}$, where $ k \in \mathbb{Z}$, the term $2Nc_1$ remains an integer, thereby preserving periodicity and preventing any additional phase distortion. Then for $\mathbf{\Lambda}r[n] = s[m]$, the scaling the time-domain signal by $2Nc_1$ gives 
\begin{equation}
    \begin{aligned}
       \mathbf{\Lambda}r[2Nc_1n] = s[2Nc_1m]. 
    \end{aligned}
\end{equation}
Hence the autocorrelation function in \eqref{equ: auto_afdm4} can be expressed as
\begin{equation}
    \begin{aligned}
    R_{rr}^a[m,\dot{m}] = R_{rr}^t[2Nc_1(m-\dot{m})] = R_{rr}^t[2Nc_1m_{\delta}].
        \label{equ: auto_afdm5}
    \end{aligned}
\end{equation}
Consequently, the autocorrelation function in the affine domain is a function of the indices difference $m_{\delta}=m-\dot{m}$. Additionally, as the variance can be calculated from the autocorrelation function for a \ac{WSS} signal with $\mathbb{E}\left\{\mathbf{\Lambda}\mathbf{r}\right\}=\mu_r^a =0$, such as
\begin{equation}
    \begin{aligned}
     VAR\left\{r[n]\right\} = R_{rr}^a[0],
        \label{equ: variance}
    \end{aligned}
\end{equation}
which is constant across the affine domain. As illustrated in Fig. \ref{fig: Affine_domain_representation} (a), for a time-domain \ac{AWGN} signal with zero mean, the corresponding signal in the affine domain also exhibits AWGN characteristics. Based on this observation, we formally state Theorem 1 for \ac{WSS} signals. \\
\textbf{Theorem 1.} \textit{For a time-domain \ac{WSS} signal $r[n]$, its representation in the affine domain remains \ac{WSS}, provided that its mean satisfies $\mu_r^t = 0$. Furthermore, it preserves AWGN characteristics, following the distribution $\sim \mathcal{N}\left(0, R_{rr}^a[0]\right)$.} \\
Note that the previous theorem holds when $2Nc_1$ is an integer and as long as the mean of the time-domain signal is zero. Next, we expand this theorem to a general \ac{OFDM} signal.
\subsection{OFDM in Affine-Domain}
Consider the \ac{OFDM} signal $\mathbf{F}^H \mathbf{x}$ transformed to the affine domain using $N$-points DAFT such as 
\begin{equation}
    \begin{aligned}
    \mathbf{y} = \boldsymbol{\Lambda}_{c_1} \mathbf{F} \boldsymbol{\Lambda}_{c_2} \mathbf{F}^H \mathbf{x},
        \label{equ: OFDM_affine}
    \end{aligned}
\end{equation}
where $\mathbf{x} =\left[x_0, \cdots, x_{N}\right]^T \in \mathbb{C}^{N\times 1}$ are drawn from a set of independent and identically distributed (i.i.d.) random QAM symbols. For $N$ asymptotically large, the real and imaginary parts of $\mathbf{x}$ follow a gaussian distribution with zero mean \cite{prasad2004ofdm}. Then, when a Gaussian coding code-book is applied we can have $\mathbb{E}\left\{\mathbf{x}\right\}=0$. Given that, the covariance matrix of $\mathbf{y}$ is given by 
\begin{equation}
    \begin{aligned}
    \mathbf{C}_y & = \mathbb{E}\left\{\mathbf{y}\mathbf{y}^H\right\} = \mathbb{E}\left\{\boldsymbol{\Lambda}\mathbf{F}^H\mathbf{x}\mathbf{x}^H\mathbf{F}\boldsymbol{\Lambda}^H\right\} \\
    &  = \boldsymbol{\Lambda}\mathbf{F}^H \mathbf{C}_x \mathbf{F}\boldsymbol{\Lambda}^H,
        \label{equ: cov_ofdm}
    \end{aligned}
\end{equation}
with $\mathbf{C}_x= \mathbb{E}\left\{\mathbf{x}\mathbf{x}^H\right\}$ given as
\begin{equation} 
    \begin{aligned}
    \mathbf{C}_x=\left[\begin{array}{cccc}
\mathbb{E}\left\{|x_0|^2\right\} & 0 & \cdots & 0 \\
0 & \mathbb{E}\left\{|x_1|^2\right\} & \cdots & 0 \\
\vdots & & & \vdots \\
0 & 0 & \cdots & \mathbb{E}\left\{|x_{N-1}|^2\right\}
\end{array}\right].
        \label{equ: cov_ofdm1}
    \end{aligned}
\end{equation}
Under the per-sample power constraint \cite{heath2018foundations}, the transmitted symbols are assumed to have uniform power, such that $\mathbb{E}\left\{|x_n|^2\right\}= \varsigma \quad \forall n$. With this assumption, the covariance matrix in \eqref{equ: cov_ofdm1} simplifies to $\mathbf{C}_x = \varsigma \mathbf{I}$. Given that $\mathbf{\Lambda}$ and $\mathbf{F}$ are unitary matrices, it follows that $\mathbf{C}_y = \mathbf{C}_x = \varsigma \mathbf{I}$, which is also a diagonal matrix. Furthermore, given that the expectation of the generated symbols satisfies $\mathbb{E}\left\{\mathbf{x}\right\} = 0$, the expectation of the affine-domain representation becomes $\mathbb{E}\left\{\mathbf{y}\right\}= \mathbf{\Lambda}\mathbf{F}^H \mathbb{E}\left\{\mathbf{x}\right\} = 0$. This implies that an \ac{OFDM} signal exhibits \ac{AWGN} characteristics in the affine domain.\\ 
When the data symbols \( x_n \) exhibit a non-uniform power distribution, the covariance matrix \( \mathbf{C}_x \) no longer satisfies \( \mathbf{C}_x = \varsigma \mathbf{I} \). As a result, the covariance matrix of the affine-domain signal, \( \mathbf{C}_y \), is no longer strictly diagonal. However, the magnitudes of the off-diagonal elements remain significantly smaller compared to the diagonal entries, ensuring that the power distribution along the main diagonal of \( \mathbf{C}_y \) remains relatively uniform. Mathematically, this can be expressed as
\begin{equation}
     \begin{aligned}
     \text{tr}(\mathbf{C}_y) \approx \sum_{n=0}^{N-1}\mathbb{E}\left\{|x_n|^2\right\} = N \bar{\varsigma},
         \label{equ: off-diagonal}
     \end{aligned}
\end{equation}
where $\bar{\varsigma}$ represents the average power per sample. The observed diagonal dominance in the covariance matrix arises due to the quadratic phase terms \( \boldsymbol{\Lambda}_{c_{\{1,2\}}} \), which introduce sufficient statistical randomization, provided that the refining coefficients \( c_{\{1,2\}} \) are appropriately chosen. \textcolor{black}{Typically, $c_1$ is selected to adapt to the channel maximum Doppler shift, whereas $c_2$ is set be a rational number sufficiently smaller than $\frac{1}{N}$ \cite{bemani2023affine}. Therefore, $c_1$ and $c_2$ are given by
    \begin{equation}
        \begin{aligned}
         c_1 = \frac{k}{2N}, \quad c_2 = \frac{1}{N^2}, \quad \text{where } k \in \mathbb{Z}.   
        \end{aligned}
    \end{equation}}  
This randomization effectively de-correlates the samples, thereby reducing the power of the off-diagonal elements. Moreover, as the number of samples \( N \) increases, the de-correlation effect becomes more pronounced, further reinforcing the approximate diagonal structure of \( \mathbf{C}_y \). Based on this observation, the following theorem is presented for an \ac{OFDM} signal. \\ 
\textbf{Theorem 2.} \textit{An OFDM signal with i.i.d. data symbols and normalized power allocation satisfying $\frac{1}{N} \sum_{n=0}^{N-1} \mathbb{E}\left\{|x_n|^2\right\} = \bar{\varsigma}$, exhibits a uniform power distribution in the affine domain and behaves as an \ac{AWGN} process, with $ \sim \mathcal{N} \left(0, \bar{\varsigma} \right)$.
}\\
Figure \ref{fig: Affine_domain_representation}(b) illustrates the behavior of an \ac{OFDM} signal (in red) within the affine domain, where the \ac{OFDM} samples exhibit a noise-like scattering across the localized \ac{AFDM} carrier. When superimposed on the localized carrier, these scattered samples degrade its power level while preserving its structural representation. Furthermore, Fig. \ref{fig: Affine_domain_representation}(c) and (d) depict the power difference between the localized \ac{AFDM} pulse and the OFDM samples in the affine domain for different \ac{OFDM} and \ac{AFDM} sizes. Specifically, for $N_r < N_c$, the resulting noisy samples possess power levels comparable to those of the \ac{AFDM} carrier sidelobes, leading to a degradation in the combined signal power. However, for $N_r > N_c$, the large-size DAFT disperses the \ac{OFDM} samples widely across the affine domain, thereby significantly reducing the interference over the \ac{AFDM} carrier. This observation is consistent with the result in \eqref{equ: off-diagonal}.

\section{Waveform-Domain SIC Algorithm}
This section introduces the waveform-domain SIC algorithm designed for the proposed IBFD-\ac{ISAC} system. The proposed scheme incorporates noise suppression in the affine domain, iterative windowing, and time-domain spreading of the SIC-processed signal. Following this, the performance of the system is evaluated based on radar-specific performance metrics. 
\subsection{Affine-Domain Cancellation}
Consider the received signal in \eqref{equ: recieved_signal_matrix}, after  applying DAFT, it results in 
\begin{equation}
    \begin{aligned}
    \bold{y}_j = \sum_{i=1}^L \hat{\alpha}_i \mathbf{\Lambda} \bold{\Pi}^{l_i} \bold{\Delta}^{\kappa_i} \hat{\bold{s}}_{j,n_{l_i}} + \beta \mathbf{\Lambda}\bold{s}^c_j + \bold{w}_j^a,
        \label{equ: y_1}
    \end{aligned}
\end{equation}
where $\bold{w}_j^a=\mathbf{\Lambda}\bold{w}_j $ denotes the \ac{AWGN} in the affine domain. Note that since $\Lambda$ is a unitary matrix, $\bold{w}_j^a$ preserves the same statistical properties as $\bold{w}_j$. For a single target, within the radar reception period, $\bold{y}_j = \mathbf{\Lambda} \bold{r}_j $ can be decomposed into
\begin{equation}
    \begin{aligned}
      \begin{cases} 
     \hat{\alpha} \mathbf{\Lambda} \bold{\Pi}^{l} \bold{\Delta}^{\kappa} \bold{s}_{j-1}^c + \bold{w}_j^a, \quad  0\leq m \leq N_g, \\[3pt]
      \hat{\alpha} \mathbf{\Lambda} \bold{\Pi}^{l} \bold{\Delta}^{\kappa} \bold{s}_{j-1}^c + \beta \mathbf{\Lambda}\bold{s}_j^c + \bold{w}_j^a, \quad  N_g< m\leq n_l-N_g, \\[3pt]
      \beta \mathbf{\Lambda}\bold{s}_j^c + \bold{w}_j^a, \quad  n_l-N_g < m\leq n_l, \\[3pt]
      \hat{\alpha} \mathbf{\Lambda} \bold{\Pi}^{l} \bold{\Delta}^{\kappa} \bold{s}_{j}^r + \beta \mathbf{\Lambda}\bold{s}_j^c + \bold{w}_j^a, \quad n_l < m\leq N_r^{tot}+n_l, \\[3pt]
      \beta \mathbf{\Lambda}\bold{s}_j^c + \bold{w}_j^a, \quad   N_r^{tot}+n_l<m\leq N_r^{tot}+n_l+N_g, \\[3pt]
      \hat{\alpha} \mathbf{\Lambda} \bold{\Pi}^{l} \bold{\Delta}^{\kappa} \bold{s}_{j}^c + \beta \mathbf{\Lambda}\bold{s}_j^c + \bold{w}_j^a, \\ \quad \quad \quad N_r^{tot}+n_l+N_g\leq m \leq N^{tot}-N_r^{tot}-N_g, \\[3pt]
      \hat{\alpha} \mathbf{\Lambda} \bold{\Pi}^{l} \bold{\Delta}^{\kappa} \bold{s}_{j}^c + \bold{w}_j^a, \quad  N^{tot}-N_r^{tot}-N_g < m \leq N^{tot}-N_r^{tot}.
     \end{cases}
        \label{equ: y_2}
    \end{aligned}
\end{equation}
Based on the above, the target’s information is contained within the interval \( n_l < m \leq N_r^{tot} + n_l \), while the remaining components of the received signal consist of a combination of \ac{OFDM} SI, noise, and communication signals from both the previous and current \ac{PRI}s, denoted as \( \bold{s}_{j-1}^c \) and \( \bold{s}_{j}^c \), respectively. As established in \textbf{Theorem 2.}, the \ac{OFDM} signal exhibits a constant power and can be modeled as \ac{AWGN} in the affine domain. Consequently, the combined effect of \( \bold{w}_j^a \) and \( \beta \mathbf{\Lambda} \bold{s}_j^c \) can be interpreted as a superposition of noise, leading to:
\begin{equation}
    \begin{aligned}
        \bold{w}_j^{a, SI} = \beta \mathbf{\Lambda}\bold{s}_j^c + \bold{w}_j^a, \textit{ where} \quad \bold{w}_j^{a, SI}  \sim \mathcal{N}\left(0, \sigma^2+\beta \right).
    \end{aligned}
\end{equation}
For the \ac{OFDM} communication signal which passes by the channel, that is $\hat{\alpha} \mathbf{\Lambda} \bold{\Pi}^{l} \bold{\Delta}^{\kappa} \bold{s}_{j-1}^c$, it can be shown for 
\begin{equation}
    \begin{aligned}
     r[m]=\sum_{k}x[k]e^{j\frac{2\pi k}{N}(n-l)} e^{j\frac{2\pi \kappa n}{N}} \Lambda(n,m),   
    \end{aligned}
\end{equation}
where $\Lambda(n,m) = \sum_{n=0}^{N-1}e^{-j2\pi(c_1n^2+\frac{nm}{N}+c_2m^2)}$, the autocorrelation function $\mathbb{E}\left\{r[m]r^*[\dot{m}]\right\}$ is given by
\begin{equation}
    \begin{aligned}
   R^a_{rr} & = \Lambda(n,m) \Lambda^*(\dot{n},\dot{m})\mathbb{E}\left\{x[n]x^*[\dot{n}]\right\} e^{j\frac{2\pi \kappa}{N}(n-\dot{n})} 
   \\& = \underbrace{\Lambda^*(\dot{n},\dot{m}) R^t_{xx}[n-\dot{n}]\Lambda(n,m)}_{\bar{R_{xx}^t}}e^{j\frac{2\pi \kappa}{N}(n-\dot{n})}.
    \label{equ: ofdm_receiced_affine_auto}
    \end{aligned}
\end{equation}
From \eqref{equ: auto_afdm5}, the expression $\bar{R_{xx}^t}=R_{xx}^t[2Nc_1(m-\dot{m})]$ holds; however, the summation over \( n \) and \( \dot{n} \) also incorporates the Doppler exponent. Notably, both the time-lag and the Doppler exponent manifest as a combined shift in the affine domain, as demonstrated in \cite{bemani2023affine}. Then,
\begin{equation}
    \begin{aligned}
   R^a_{rr} &= \Lambda^*(\dot{n},\dot{m}) R^t_{xx}[m-(2Nc_1+\kappa)\dot{n}]e^{-j\frac{2\pi \kappa}{N}\dot{n}}\\
   & =  R^t_{xx}[(2Nc_1+\kappa)m_{\delta}].
    \label{equ: ofdm_receiced_affine_auto}
    \end{aligned}
\end{equation}
Consequently, it can be deduced that the received \ac{OFDM} signal, after propagating through the channel, exhibits characteristics similar to those of \ac{AWGN}, maintaining a uniform power distribution in the affine domain. This behavior is illustrated in Fig. \ref{fig: Affine_domain_representation}(e), where the \ac{OFDM} signal retains its \ac{AWGN}-like properties even after passing through a two-tap channel. However, due to channel variations, the resulting phase distortions disrupt the uniform \ac{SNR} distribution in the affine domain, leading to an increase in the sidelobe levels of the combined signal, which needs adequate suppression. 
\par Following that, \eqref{equ: y_2} simplifies to
\begin{equation}
 \begin{aligned}
 \begin{cases} 
      \bold{w}_j^{tot}, \quad  0\leq m \leq n_l,  \\[3pt]
      \hat{\alpha} \mathbf{\Lambda} \bold{\Pi}^{l} \bold{\Delta}^{\kappa} \bold{s}_{j}^r + \bold{w}_j^{a, SI}, \quad n_l < m\leq N_r^{tot}+n_l, \\[3pt]
       \bold{w}_j^{tot},\quad   N_r^{tot}+n_l < m \leq N^{tot}-N_r^{tot},
     \end{cases}
     \label{equ: y_3}
 \end{aligned}   
\end{equation}
where $\mathbf{w}_j^{tot} \sim \mathcal{N}\left(0, \sigma_{tot}\right)$  with $\sigma_{tot}$ denoting the total noise variance of the \ac{SI}, AWGN, and the received \ac{OFDM} signal in the affine domain. Theoretically, since the transmitted communication frame is known to the radar receiver, the resulting SI can be perfectly eliminated by subtracting \( \beta \mathbf{\Lambda} \mathbf{s}^c_j \) from \( \mathbf{y}_j \). However, in practice, non-idealities in the receiver hardware, including non-linearities in the low-noise amplifier, the limited dynamic range of the ADC, and quantization noise, introduce residual distortions. As a result, residual SI persists even after the SIC process. Following the removal of the SI component, the resulting signal can be expressed as:
\begin{equation}
    \begin{aligned}
    \Bar{\bold{y}}_j = \sum_{i=1}^L\hat{\alpha}_i \mathbf{\Lambda} \bold{\Pi}^{l_i} \bold{\Delta}^{\kappa_i} \hat{\bold{s}}_{j,n_{l_i}} + \varepsilon \mathbf{\Lambda}\bold{s}^c_j + \bold{w}_j^a.
        \label{equ: y_4}
    \end{aligned}
\end{equation}
\begin{algorithm}[t]
    \caption{The Proposed Iterative Windowing Algorithm.}
    \SetKwInOut{Input}{Input}
    \SetKwInOut{Output}{Output}
    \SetKwInOut{KwInitialization}{Initialization}
    \DontPrintSemicolon
    \Input{$\bold{y}_j, \bold{w}, L, \beta \bold{s}_j^c, \zeta_{iter}, N_w$.}
    \Output{$\Bar{\bold{y}}_j^{w_{\rho}}$, $\left\{m^b_{iter_{\rho}}\right\}_{b=1}^{Q_{\rho}}$, $\rho$.}
    \KwInitialization{$\Bar{\bold{y}}_j=\bold{y}_j-\beta \mathbf{\Lambda}\bold{s}_j^c$, $\left\{m^b_{iter_1}\right\}_{b=1}^{Q_1} = \emptyset$, $Q_1=0$.}

    \SetKwFunction{Fone}{InitialPeaksSelection}
    \SetKwFunction{Ftwo}{IterativeWindowing}

    \SetKwProg{Fn}{Function}{}{end}

    \Fn{\Fone}{        
        \For{$m = 0: (N^{tot}-N^{tot}_{r})-1$}{        
            \eIf{$\left|\Bar{y}_j[m]\right|\geq \zeta_{iter_1}$}{
                $ m \in \left\{m^b_{iter_1}\right\}_{b=1}^{Q_1}, \quad Q_1 = Q_1+1$
            }{}
        }
    }

    Calculate $\Bar{\bold{y}}_j^{w_1}$ as in \eqref{equ: windowing}, limit search range to $\quad m^{1}_{iter_1}-\frac{N_w}{2} \leq m \leq m^{Q_1}_{iter_1}+\frac{N_w}{2}$\\

    \Fn{\Ftwo}{
        \While{$\left|\left\{m^b_{iter_{u}}\right\}\right|>\left|\left\{m^b_{iter_{u-1}}\right\}\right|$}{
            \For{$m = 0: N_{w}+m^{Q_1}_{iter_1}-m^{1}_{iter_1}$}{        
                \eIf{$\left|\Bar{y}_j^{w_u}[m]\right|\geq \zeta_{iter_u}$}{
                    $ m \in \left\{m^b_{iter_u}\right\}_{b=1}^{Q_u}, \quad Q_u = Q_u+1$
                }{}
            }
            $u = u+1$, $\zeta_{iter_{u}}\rightarrow \zeta_{iter_{u+1}}$,\\
            Calculate $\Bar{\bold{y}}_j^{w_{u}}$.
        }
    }
    \label{alg: 1}
\end{algorithm}
\begin{figure*}[!t]
\normalsize
\begin{equation}\tag{48}
\begin{aligned}
       \bar{P}_{j,zp}^{w_{\rho , \kappa} }[p,q] &= \sum^{L}_{i=1}\hat{\alpha}_i N_r^px_r^p 
e^{j\frac{2\pi}{N}\left(N_sc_1\left(p-l_i\right)^2-\kappa_ip+N_sc_2m^2+m(p-l_i)\right)}\sum^{N_s-1}_{a=0} \Upsilon^{l_i}_{N_{z_p}} \big[\dot{q}a \big] \Upsilon^{\kappa_i}_{N_{z_p}}\big[\dot{p}a\big]+\hat{\bold{w}}_j^{tot}.
      \label{eq: zp_k}
\end{aligned}
\end{equation}
\hrulefill
\vspace*{0.5pt}
\begin{equation}\tag{49}
\begin{aligned}
        \bar{P}_{j,zp}^{w_{\rho , l}}[p,q]  &= \sum^{L}_{i=1}\hat{\alpha}_i N_r^px_r^p
e^{j\frac{2\pi}{N}\left(N_sc_1\left(p-l_i\right)^2-\kappa_ip+N_sc_2m^2+m(p-l_i)\right)}\sum^{N_{s}-1}_{a=0} \Upsilon^{l_i}_{N_{z_p}} \big[\ddot{p}a \big] \Upsilon^{\kappa_i}_{N_{z_p}}\big[\ddot{q}a\big]+\hat{\bold{w}}_j^{tot}.
      \label{equ: zp_l}
\end{aligned}
\end{equation}
 \hrulefill
\vspace*{0.5pt}
\end{figure*}
Here, \( \varepsilon \) denotes the magnitude of the residual SI. This residual SI, combined with the high sidelobe levels of the communication signal after propagating through the channel, introduces significant distortion. Such distortion must be efficiently mitigated to ensure accurate target detection through a localized windowing process. Consider a threshold-based scheme, where the indices \( \left\{m^b_{iter_1}\right\}_{b=1}^{Q_1} \) that exceed an initial predefined threshold \( \zeta_{iter_1} \in \left\{\zeta_{iter_u}\right\}_{u=1}^{\rho} \) within \( \Bar{\bold{y}}_j \) are selected for further processing. Specifically, an element-wise multiplication windowing operation \( \bold{w} \) is applied, with a window size of \( N_w = 2N_r^{tot} \) centered around each detected peak $m^b_{iter_1}$, where \( \rho \) denotes the total number of iterations.  This can be expressed as, 
\begin{equation}\small
    \begin{aligned}
    \Bar{\bold{y}}_j^{w_1} =   \sum_{b=1}^{Q_1}\sum_{i=1}^L & \mathbbm{1}_{m_{w_1}^b}\Bar{\bold{y}}_j \odot \bold{w} = \sum_{b=1}^{Q_1} \sum_{i=1}^L \hat{\alpha}_i (\mathbbm{1}_{m_{w_1}^b} \mathbf{\Lambda} \bold{\Pi}^{l_i} \bold{\Delta}^{\kappa_i} \bold{s}_{j}^r) \odot \bold{w} \\
    & + (\mathbbm{1}_{m_{w_1}^b} \varepsilon \mathbf{\Lambda}\bold{s}_j^c + \mathbbm{1}_{m_{w_1}^b} \bold{w}_j^a) \odot \bold{w}\\
    & \textit{ for,} \quad m^{1}_{iter_1}-\frac{N_w}{2} \leq m \leq m^{Q_1}_{iter_1}+\frac{N_w}{2},
        \label{equ: windowing}
    \end{aligned}
\end{equation}
given that,
\begin{equation}
    \begin{aligned}
      \mathbbm{1}_{m_{w_1}^b} = \begin{cases}
          1, \quad m \in m_{w_1}^b \triangleq [m^b_{iter_1}-\frac{N_w}{2},m^{b}_{iter_1}+\frac{N_w}{2}] \\
          0, \quad m \notin m_{w_1}^b
      \end{cases}.  
    \end{aligned}
\end{equation}
Subsequently, the iterative windowing scheme will lock solely on the samples within the range \( m^{1}_{iter_1} - \frac{N_w}{2} \leq m \leq m^{Q_1}_{iter_1} + \frac{N_w}{2} \), where the radar information is concentrated. By enlarging \( N_w \), the initial windowing process ensures that all targets are captured. In the subsequent iterations, the window size \( \bold{w}^u \) is reduced, becoming limited to adjacent carriers, i.e., \( N_w^{iter} = N_r^{tot} \). As the windowing process effectively suppresses noise, additional peaks \( \left\{ m^b_{iter_u} \right\}_{b=1}^{Q_u} \) may emerge in the \( u \)-th iterations. Fig. \ref{fig: Affine_domain_representation}(f) demonstrates that the windowing process applied at the localized carrier effectively diminishes the sidelobes while simultaneously magnifying weak targets that were once concealed by elevated SI residual levels during the initial cancellation phase. In the following iterations, the same peak detection and windowing process continues with a reduced threshold \( \zeta_{iter_u} \), until the total number of detected peaks maintains throughout two consecutive iterations, such as
\begin{equation}
    \begin{aligned}
   & \Bar{\bold{y}}_j^{w_u}  =   \sum_{b=1}^{Q_u}\sum_{i=1}^L \mathbbm{1}_{m_{w_u}^b}\Bar{\bold{y}}_j^{w_{u-1}} \odot \bold{w}^u\\
    & \textit{ for}, \quad \left|\left\{m^b_{iter_{u}}\right\}_{b=1}^{Q_{u}}\right|>\left|\left\{m^b_{iter_{u-1}}\right\}_{b=1}^{Q_{u-1}}\right|, u = 1, \cdots, \rho.
        \label{equ: iterations}
    \end{aligned}
\end{equation}
The resulting signal after $\rho$ iterations can be written as
\begin{equation}
    \begin{aligned}
    \Bar{\bold{y}}_j^{w_{\rho}}& =  \sum_{b=1}^{Q_{\rho}} \sum_{i=1}^L\hat{\alpha}_i (\mathbbm{1}_{m_{w_{\rho}}^b} \mathbf{\Lambda} \bold{\Pi}^{l_i} \bold{\Delta}^{\kappa_i} \bold{s}_{j}^r) \odot \bold{w}^{{\rho}} + \varepsilon_{\rho} \bold{w}_j^{{a, SI}_\rho}  \\
    & \textit{ for,} \quad m^{1}_{iter_{1}}-\frac{N_w}{2} \leq m \leq m^{Q_{1}}_{iter_{1}}+\frac{N_w}{2},
        \label{equ: final_y}
    \end{aligned}
\end{equation}
where $m_{w_{\rho}}^b \in [m^b_{iter_{\rho}}-\frac{N_w}{2},m^{b}_{iter_{\rho}}+\frac{N_w}{2}]$, and 
\begin{equation}
    \begin{aligned}
      \bold{w}_j^{{a, SI}_\rho} = (\mathbbm{1}_{m_{w_{\rho}}^b}\varepsilon \mathbf{\Lambda}\bold{s}_j^c + \mathbbm{1}_{m_{w_{\rho}}^b} \bold{w}_j^a) \odot \bold{w}^{{\rho}}.  
    \end{aligned}
\end{equation}
Here, \( \varepsilon_{\rho} \) represents the magnitude of the residual SI after successive windowing. It is important to note that \( \bold{w}_j^{{a, SI}_\rho} \) remains a zero-mean \ac{AWGN} with its power level directly proportional to the residual SI magnitude \( \varepsilon_{\rho} \), distributed across the samples of \( \Bar{\bold{y}}_j^{w_{\rho}} \). The proposed affine-domain iterative windowing process is outlined in \textbf{Algorithm 1}.
\subsection{Time-Domain Spreading}
After performing the iterative windowing, the range and velocity of the targets can be extracted from the resulting signal \( \Bar{\bold{y}}_j^{w_{\rho}} \) in the time domain. Specifically, the \ac{CPI} matrix is constructed by accumulating the signals \( \Bar{\bold{y}}_j^{w_{\rho}} \) from each \ac{PRI} over a total of \( J \) \ac{PRI}s. This results in the matrix \( \Bar{\bold{Y}}= [\Bar{\bold{y}}_1^{w_{\rho}}, \cdots, \Bar{\bold{y}}_j^{w_{\rho}}, \cdots, \Bar{\bold{y}}_J^{w_{\rho}}] \in \mathbb{C}^{N_s\times J} \), where \( N_s = N_{w} + m^{Q_1}_{iter_1} - m^{1}_{iter_1} \). Subsequently, the matrix \( \bar{\bold{Y}} \) is transformed into the time domain by applying an IDAFT to each \( \Bar{\bold{y}}_j^{w_{\rho}} \), yielding \( \Bar{\bold{r}}_j^{w_{\rho}} = \bold{\Lambda}^H\Bar{\bold{y}}_j^{w_{\rho}} \). This results in the time-domain SI-free matrix \( \bar{\bold{R}}= [\Bar{\bold{r}}_1^{w_{\rho}}, \cdots, \Bar{\bold{r}}_j^{w_{\rho}}, \cdots, \Bar{\bold{r}}_J^{w_{\rho}}] \in \mathbb{C}^{N_s\times J} \). However, rather than processing the entire matrix \( \bar{\bold{R}} \) as in conventional radar CPI processing, the target information can be extracted directly from each vector \( \Bar{\bold{r}}_j^{w_{\rho}} \) by leveraging the chirp-carrier behavior in the PCTD, as detailed in \cite{arousnovel}. In this domain, the delay and Doppler components become decoupled and manifest as distinct shifts. This property holds even though the Doppler phase term \( e^{j2\pi \nu t} \) diminishes for \( \nu T_r^{tot} \ll 1 \) due to Doppler spreading in the PCTD. Consequently, Doppler estimation can be performed without requiring the coherent integration gain typically associated with the slow-time processing of the CPI. Following that, for Doppler estimation, the vector $\Bar{\bold{r}}_j^{w_{\rho}}$ is reshaped and zero-padded such as
\begin{equation}
    \begin{aligned}
    \Bar{\bold{r}}_j^{w_{\rho, \kappa}} \rightarrow \Bar{\bold{R}}_j^{w_{\rho, \kappa}}, \quad \Bar{\bold{R}}_j^{w_{\rho, \kappa}} \rightarrow \Bar{\bold{R}}_{j,zp}^{w_{\rho , \kappa}},  
        \label{equ: radar_proc_1}
    \end{aligned}
\end{equation}
where $\Bar{\bold{R}}_{j,zp}^{w_{\rho , k}}$ is the Doppler zero-padded reshaped matrix, for $\sqrt{N_s} \in \mathbb{Z}^+$ being even, it can be written as
\begin{equation}
    \begin{aligned}
    \Bar{R}_{j,zp}^{w_{\rho , \kappa}}[p, a] = \begin{cases}
        \Bar{R}_{j}^{w_{\rho , \kappa}}[p, a], \text{ for} \quad 0 \leq p, a \leq \sqrt{N_s}\\
        0 , \text{ for} \quad \sqrt{N_s} < p, a \leq N_{z_p}
    \end{cases}   ,
        \label{equ: reshaped}
    \end{aligned}
\end{equation}
where $N_{zp} = z_p \sqrt{N_s}$ and $z_p$ is the zero-padding factor. Applying IDFT over the columns of $\Bar{\bold{R}}_{j,zp}^{w_{\rho , \kappa}}$ gives the PCTD representation of the signal, given by $\Bar{\bold{P}}_{j,zp}^{w_{\rho , k}} = \bold{F}^{H}\Bar{\bold{R}}_{j,zp}^{w_{\rho , \kappa}}$. Note that the zero-padding will not only enable Doppler detection within a single \ac{PRI}, but also spreads the residual \ac{SI} across the matrix $\Bar{\bold{P}}_{j,zp}^{w_{\rho , \kappa}}$ thereby further reducing its effect on target detection. For range detection, the delay shift in the time domain is transformed into phase variations by applying a DFT matrix to the vector $\Bar{\bold{r}}_j^{w_{\rho}}$ such as $\Bar{\bold{r}}_j^{w_{\rho}, l}=\bold{F} \Bar{\bold{r}}_j^{w_{\rho}}$. Subsequently, the same process of zero-padding and IDFT spreading is applied, leading to 
\begin{equation}
    \begin{aligned}
    \Bar{\bold{r}}_j^{w_{\rho, l}} \rightarrow \Bar{\bold{R}}_j^{w_{\rho, l}}, \quad \Bar{\bold{R}}_j^{w_{\rho, l}} \rightarrow \Bar{\bold{R}}_{j,zp}^{w_{\rho , l}}.   
        \label{equ: radar_proc_1}
    \end{aligned}
\end{equation}
Performing IDFT gives $\Bar{\bold{P}}_{j,zp}^{w_{\rho , l}} = \bold{F}^{H}\Bar{\bold{R}}_{j,zp}^{w_{\rho , l}}$. The resulting PCTD representation of the Doppler and delay matrices $\Bar{\bold{P}}_{j,zp}^{w_{\rho , \kappa}}$ and $\Bar{\bold{P}}_{j,zp}^{w_{\rho , l}}$ formulas are given in \eqref{eq: zp_k} and \eqref{equ: zp_l}, respectively, where $x_r^p$ represents the pilots carriers for each \ac{AFDM} symbol. $ \Upsilon^{l_i}_{N_{z_p}}$ and $ \Upsilon^{\kappa_i}_{N_{z_p}}$ contain the decoupled delay and Doppler shifts, for the Doppler matrix $P_{j,zp}^{w_{\rho , \kappa} }$ they are given by 
\begin{equation}\setcounter{equation}{50}
    \Upsilon^{l_i}_{N_{z_p}}\big[\dot{q} a\big] = \begin{cases} [l_i]_1 =0, &  \begin{cases} N_{z_p}, & [q-l_i+\frac{\sqrt{N_s}}{2}]_{\sqrt{N_s}}=0 \\ 0 & \textit{otherwise } \end{cases}\\ [l_i]_1 \neq 0, & \frac{e^{j2\pi(q-l_i+\frac{\sqrt{N_s}}{2})}-1}{e^{j\frac{2\pi}{\sqrt{N_s}}(q-l_i+\frac{\sqrt{N_s}}{2})}-1}
    \end{cases},
    \label{equ: sinc_delay}
\end{equation}
\begin{equation}
    \Upsilon^{\kappa_i}_{N_{z_p}}\big[\dot{p}a\big] = \begin{cases} [z_p\kappa_i]_1 =0, &  \begin{cases} N_{z_p}, & [p+z_p\kappa_i]_{\sqrt{N_s}}=0 \\ 0 & \textit{otherwise } \end{cases}\\ [z_p\kappa_i]_1 \neq 0, & \frac{e^{j2\pi(p+z_p\kappa_i)}-1}{e^{j\frac{2\pi}{N_s}(p+z_p\kappa_i)}-1}
    \end{cases},
    \label{equ: sinc_Doppler}
\end{equation}
with $\dot{q} =q-l_i+\frac{\sqrt{N_s}}{2}$ and $\dot{p} = p+z_p\kappa_i$. Whereas for $P_{j,zp}^{w_{\rho , l} }$ we have
\begin{equation}\small          \Upsilon^{l_i}_{N_{z_p}}\big[\ddot{p}a\big] = \begin{cases} [z_pl_i]_1 =0, &  \begin{cases} N_{z_p}, & [p-z_pl_i+\frac{\sqrt{N_s}}{2}]_{\sqrt{N_s}} =0 \\ 0, & \textit{otherwise } \end{cases}\\ [z_pl_i]_1 \neq 0, & \frac{e^{j2\pi(p-z_pl_i+\frac{\sqrt{N_s}}{2})}-1}{e^{j\frac{2\pi}{\sqrt{N_s}}(p-z_pl_i+\frac{\sqrt{N_s}}{2})}-1}
    \end{cases},
    \label{equ: sinc_delay2}
\end{equation}
\begin{equation}
    \Upsilon^{\kappa_i}_{N_{z_p}}\big[\ddot{q}a\big] = \begin{cases} [\kappa_i]_1 =0, &  \begin{cases} N_{z_p}, & [q+\kappa_i]_{\sqrt{N_s}}=0 \\ 0 & \textit{otherwise } \end{cases}\\ [\kappa_i]_1 \neq 0, & \frac{e^{j2\pi(q+\kappa_i)}-1}{e^{j\frac{2\pi}{N_s}(q+\kappa_i)}-1}
    \end{cases},
    \label{equ: sinc_Doppler2}
\end{equation}
with  $\ddot{p} = p-z_pl_i+\frac{\sqrt{N_s}}{2}$ and $\ddot{q} =q+\kappa_i$. 
\par From \eqref{eq: zp_k} and \eqref{equ: zp_l}, a higher $N_r^p$ leads  to more localized pulses in the PCTD and hence better target extraction. Moreover, the residual SI, noise, and other remaining components are encapsulated within \( \hat{\bold{w}}_j^{tot} \), which is dispersed across the columns of \( \Bar{\bold{P}}_{j,zp}^{w_{\rho , \kappa}} \) and \( \Bar{\bold{P}}_{j,zp}^{w_{\rho , l}} \). Additionally, due to the iterative windowing process in the affine domain, time-domain detection can be directly executed without requiring additional digital cancellation. This detection is achieved via any threshold-based technique, mapping the corresponding delay and Doppler indices as defined in \eqref{equ: sinc_delay2} and \eqref{equ: sinc_Doppler}.
\subsection{Performance Analysis}
In this subsection, the main performance metrics of the proposed IBFD-\ac{ISAC} system are analyzed. 
\subsubsection{SINR}
The explicit expression of the SIC signal in time-domain before converting to the PCTD can be given by
\begin{equation}
    \begin{aligned}
     \Bar{\bold{r}}_j^{w_{\rho}} =  \left(\sum_{i=1}^L \hat{\alpha}_i \boldsymbol{\Pi}^{l_i} \mathbf{\Delta}^{\kappa_i} \mathbf{s}_j^r+\varepsilon_\rho \varepsilon \mathbf{s}_j^c+\varepsilon_\rho \mathbf{w}_j\right) \odot \mathbf{w}^\rho.
        \label{equ: r_rho}
    \end{aligned}
\end{equation}
Consider $P_c$ and $P_r$ to be the total communication and radar power, respectively. Then for a normalized power over a symbol duration, we have
\begin{equation}
    \begin{aligned}
    \mathbb{E}&\left[|\Bar{\bold{r}}_j^{w_{\rho}}|^2\right]  = \mathbb{E}\overbrace{\left[\left|   \sum_{i=1}^L  \hat{\alpha}_i \sqrt{P_r T_r^{\eta}} \boldsymbol{\Pi}^{l_i} \mathbf{\Delta}^{\kappa_i} \mathbf{s}_j^r \odot \mathbf{w}^\rho \right|^2\right]}^{ \gamma_j^r} \\
    &+ \mathbb{E}\underbrace{\left[\left|  \sqrt{P_c T_c^{\eta}} \varepsilon_\rho \varepsilon \mathbf{s}_j^c \odot \mathbf{w}^\rho \right|^2\right]}_{ \gamma_j^{SI}} + 
    \mathbb{E}\underbrace{\left[\left| \varepsilon_\rho \mathbf{w}_j \odot \mathbf{w}^\rho \right|^2\right]}_{ \gamma_j^n}, 
        \label{equ: r_rho1}
    \end{aligned}
\end{equation} 
where $\gamma_j^r$, $\gamma_j^{SI}$ and $\gamma_j^n$ denote the radar signal power, the residual \ac{SI} and noise powers after SIC, respectively. Simplifying $\gamma_j^r$ gives 
\begin{equation}
    \begin{aligned}
    \gamma_j^r = \sum_{i=1}^L \left| \hat{\alpha}_i \right|^2  P_r T_r^{\eta} \sum_{n=0}^{N_s-1} |s_j^r[n]|^2  |w^\rho[n]|^2. 
        \label{equ: r_rho2}
    \end{aligned}
\end{equation} 
The samples $|w^\rho[n]|^2$ decay at each windowing iteration $u$, where $|w^u[n]|<1 \quad \forall n$, hence $C_{\rho} = \sum_{n=0}^{N_s-1} |w^\rho[n]|^2 <1$. Moreover, for large $N_s$, $\sum_{n=0}^{N_s-1} |s_j^r[n]|^2 \approx N_s$ as the samples of $s_j^r[n]$ are also uncorrelated i.i.d normalized random QAM symbols, then
\begin{equation}
    \begin{aligned}
    \gamma_j^r = \sum_{i=1}^L \left| \hat{\alpha}_i \right|^2  N_s P_r T_r^{\eta} C_{\rho}.  
        \label{equ: r_rho3}
    \end{aligned}
\end{equation} 
Similarly, simplifying $\gamma_j^c$ and $\gamma_j^n$ gives 
\begin{subequations}
    \begin{align}
        \gamma_j^{SI} &=  (\varepsilon_\rho \varepsilon)^2 N_s P_c T_c^{\eta} C_{\rho} ,
        \label{equ: r_rho4} \\
        \gamma_j^n &=  \varepsilon_\rho^2 N_0 C_{\rho} ,
        \label{equ: r_rho5}
    \end{align}
\end{subequations}
where $N_0$ is the noise power. Then, the \ac{SINR} for the $j$-th \ac{PRI} can be expressed as
\begin{equation}
    \begin{aligned}
        \gamma_j^{\text{SINR}} = \frac{\gamma_j^r}{\gamma_j^{SI}+\gamma_j^n} = \frac{1}{\varepsilon_\rho^2}  \frac{\sum_{i=1}^L \left| \hat{\alpha}_i \right|^2  N_s N_r P_r }{\varepsilon^2 N_sN_c P_c + N_0B}.
        \label{equ: SINR}
    \end{aligned}
\end{equation}
From \eqref{equ: SINR}, the \ac{SINR} of each \ac{PRI} depends on the residual windowing suppression factor \( \varepsilon_\rho^2 \). As a result, increasing the number of windowing iterations enhances \( \gamma_j^{\text{SINR}} \) by gradually reducing the residual SI. However, this improvement comes at the cost of increased computational complexity, as a higher number of iterations $\rho$ requires additional processing steps. 
\subsubsection{Probability of Detection}
Using a binary hypothesis testing problem formulation, it can be written
\begin{equation}
    \begin{aligned}
    \mathcal{H}_1: \Bar{\bold{r}}_j^{w_{\rho}}|_{n_{i}} =\bar{S}+\bar{I}, \quad \mathcal{H}_0: \Bar{\bold{r}}_j^{w_{\rho}}|_{n_{i}}=\bar{I},
        \label{equ: hypothesis}
    \end{aligned}
\end{equation}
with, 
\begin{subequations}
    \begin{align}
        \bar{I} &= \left((\varepsilon_\rho \varepsilon \mathbf{s}_j^c+\varepsilon_\rho \mathbf{w}_j) \odot \mathbf{w}^\rho \right)_{n_{i}}, \label{eq:a} \\
        \bar{S} &= \left(\sum_{i=1}^L \hat{\alpha}_i \boldsymbol{\Pi}^{l_i} \mathbf{\Delta}^{\kappa_i} \mathbf{s}_j^r \odot \mathbf{w}^\rho \right)_{n_{i}}, \label{eq:b}
    \end{align}
\end{subequations}
where $\Bar{\bold{r}}_j^{w_{\rho}}|_{n_{i}}$, $\bar{S}$ and $\bar{I}$ denote the post SIC signal, the target signal, and the combined residual \ac{SI} plus noise value at the target bin $n_{i} \in \left\{n_{1}, \cdots, n_L \right\}$, respectively. Using a threshold detector, we can write
\begin{equation}
    \begin{aligned}
        z= |\Bar{r}_j^{w_{\rho}}[n_i]|  \underset{\mathcal{H}_0}{\stackrel{\mathcal{H}_1}{\gtrless}} \zeta.
    \end{aligned}
\end{equation}
Since $\bar{I}$ is \ac{AWGN} with $\bar{I} \sim \mathcal{N}\left(0, \sigma_{\bar{I}}\right)$, then for the hypothesis $\mathcal{H}_0$, $z$ is Rayleigh distributed and its probability density function (PDF) is given by 
\begin{equation}
    \begin{aligned}
        p\left(z \mid \mathcal{H}_0\right) =\frac{2 z}{\sigma_{\bar{I}}^2} \exp \left(-\frac{z^2}{\sigma_{\bar{I}}^2}\right), \quad z>0,
        \label{equ: Rayleigh}
    \end{aligned}
\end{equation}
with the probability of false alarm given as
\begin{equation}
    \begin{aligned}
        P_{FA}=\int_{\zeta}^{+\infty} p\left(z \mid \mathcal{H}_0\right) \mathrm{d} z=\exp \left(-\frac{\zeta}{\sigma_{\bar{I}}^2}\right).
        \label{equ: PoFA}
    \end{aligned}
\end{equation}
Therefore, for $P_{FA}$ being a constant, the threshold can be be written as $\zeta =\sigma_{\bar{I}}\sqrt{-\text{ln} P_{FA}}$. For the hypothesis $\mathcal{H}_1$ corresponding to a target, $z$ follows a Rician distribution with a PDF 
\begin{equation}
    \begin{aligned}
        p\left(z \mid \mathcal{H}_1\right) =\frac{2 z}{\sigma_{\bar{I}}^2} \exp \left(-\frac{z^2+P_{\bar{S}}^2}{\sigma_{\bar{I}}^2}\right) I_0\left(\frac{2 z P_{\bar{S}}^2 }{\sigma_{\bar{I}}}\right), \quad z > 0
        \label{equ: Rician}
    \end{aligned}
\end{equation}
where $P_{\bar{S}}^2 = \left| \hat{\alpha} \right|^2  N_s P_r T_r^{\eta} C_{\rho} $, and $I_0(\cdot)$ is the modified Bessel function of the first kind. Then, the probability of detection is given as
\begin{equation}
    \begin{aligned}
        P_{D}=\int_{\zeta}^{+\infty} p_z\left(z \mid \mathcal{H}_1\right) \mathrm{d} z=Q_M\left(\sqrt{\frac{2 P_{\bar{S}}^2}{\sigma_{\bar{I}}^2}}, \sqrt{\frac{2 \zeta^2}{\sigma_{\bar{I}}^2}}\right),
        \label{equ: PoD}
    \end{aligned}
\end{equation}
where $Q_M(\cdot,\cdot)$ is  the first-order Marcum $Q$-function. By observing \eqref{equ: SINR} and \eqref{equ: PoFA}, it is evident that for the $j$-th \ac{PRI} the probability of detection simplifies to
\begin{equation}
    \begin{aligned}
        P_{D}&=Q_M\left(\sqrt{2 \gamma_j^{\text{SINR}}},\sqrt{-2\text{ln} P_{FA}}\right).
        \label{equ: PoD_final}
    \end{aligned}
\end{equation}
\subsubsection{Computational Complexity}

\begin{table}[t]
\caption{Simulation Parameters}
\begin{center}
\label{table: tbb}
\begin{tabular}{|c|c|c|c|c|}
\hline Parameter & Symbol  \\ 
\hline \textcolor{black}{SI channel model} & \textcolor{black}{Rician channel}  \\ 
\hline \textcolor{black}{Communication channel model} & \textcolor{black}{3GPP TDL channel} \\ 
\hline \textcolor{black}{Sensing channel model} & \textcolor{black}{3GPP EVA channel} \\ 
\hline Radar subcarriers & $N_r = 128$  \\ 
\hline Radar symbols & $M_r = 32$  \\ 
\hline Communication subcarriers & $N_c = 512$ \\ 
\hline Communication symbols & $M_c = 128$ \\ 
\hline Guard size & $N_g = 32$ \\ 
\hline Refining parameters & $c_1 = \frac{3}{N_r}, \quad c_2 = \frac{1}{N_r^2}$   \\ 
\hline Bandwidth & $B = 120$ MHz  \\ 
\hline Carrier frequency  & $f_c = 5.8$ GHz  \\ 
\hline  RCS &  $\sigma = 0.7$ $\text{m}^2$   \\ 
\hline Probability of false alarm  & $P_{FA} = 10^{-6}$  \\ 
\hline  Noise PSD &  $N_0 = -174$ dBmW/Hz   \\ 
\hline  Radar Tx and Rx antenna gains  &  $G_t = G_r = 18$ dBi \\   
\hline
\end{tabular}
\end{center}
\end{table}
    
In this part, the interest is to analyze the complexity of the proposed waveform-domain SIC approach in terms of the mathematical complex multiplications for each of the following steps:\\
\textbf{Affine-domain conversion:} After receiving the echo signal, it is transformed into the affine domain. This transformation involves diagonal matrices $\boldsymbol{\Lambda}_{c_1} \text{ and } \boldsymbol{\Lambda}_{c_2}$, each requiring $N_{c,g} = N^{tot}_c+2N_g$ multiplications. Applying the DFT matrix $ \mathbf{F}$ costs a complexity of $\mathcal{O}(N_{c,g} \text{log}_2(N_{c,g}))$. For large values of $N_{c,g}$, the computational cost of the two diagonal matrices, $2\mathcal{O}(N_{c,g})$, becomes negligible compared to the DFT complexity. Hence, the overall computational complexity simplifies to $\mathcal{O}_{1}=\mathcal{O}(N_{c,g} \text{log}_2(N_{c,g}))$.     \\
\textbf{Iterative windowing:} The first step defines $Q_1$ indices where an element-wise multiplication windowing process of size $N_w$ is applied, leading to $\mathcal{O}(N_{w}Q_1)$. Subsequently, for each of the next iterations $u = 2, \cdots, \rho$, the same windowing process is repeated with reduced size of $N_r^{tot}$ at each set of indices $Q_u$ such that $\mathcal{O}_2=\mathcal{O}(N_r^{tot}\sum_{u=2}^{\rho}Q_u)$.\\
\textbf{Time-domain processing:} Converting $\Bar{\bold{y}}_j^{w_{\rho}}$ to the time domain requires $\mathcal{O}(N_s \text{log}_2(N_s))$ operations, then applying $N_{zp}$ IDFT for the Doppler extraction requires $\mathcal{O}(N_{zp} \text{log}_2(N_{zp}))$ additional operations. For the delay processing, first an $N_s$-points DFT is applied then an $N_{zp}$ IDFT is performed after zero-padding the signal. Hence the total operations of this step is given by
\begin{equation}
    \begin{aligned}
    \mathcal{O}_3 = 2\mathcal{O}\left(N_s \text{log}_2(N_s)\right)+2\mathcal{O}\left(N_{zp} \text{log}_2(N_{zp})\right).
        \label{equ: cc_1}
    \end{aligned}
\end{equation}
Finally, the overall computational complexity  of the proposed SIC method can be given by
\begin{equation}
    \begin{aligned}
    \mathcal{O}_{tot} & = \mathcal{O}(N_{c,g} \text{log}_2(N_{c,g}))+ \mathcal{O}(N_r^{tot}\sum_{u=2}^{\rho}Q_u)+ \mathcal{O}(N_w Q_1)\\
    &  + 2\mathcal{O}\left(N_s \text{log}_2(N_s)\right)+2\mathcal{O}\left(N_{zp} \text{log}_2(N_{zp})\right).
        \label{equ: cc_2}
    \end{aligned}
\end{equation}
Note that the element-wise multiplication operations can be neglected for small $\rho$ and $Q_u$. Consequently, the system conserves a relatively low complexity in practical considerations for multicarrier radar systems.

\subsubsection{Spectral Efficiency}
The proposed system primarily employs \ac{OFDM} for data transmission, while adopting a similar transmission scheme for \ac{AFDM} at reduced data rates. Since the latter is used for sensing applications, the number of populated data carriers is lower compared to \ac{OFDM} signal. Hence, depending on the desired sensing performance, the data rates of the \ac{AFDM} signal are dynamically adjusted. Specifically, to achieve higher \ac{SINR} values, a greater number of high-power pilot carriers is allocated to each \ac{AFDM} symbol. Furthermore, to conserve the chirp-carriers representation in the PCTD, the \ac{AFDM} pilot-carriers in the affine domain must have equidistant spacing relative to the remaining data carriers, such as    

\begin{equation}
    \begin{aligned}
        \mathbf{x}_r = \begin{cases}
   x_r^p, & m=m_p \quad \quad p = 1, \ldots, N_r^p \\ 
   x_r^d, & \textit{otherwise }\end{cases}.
      \label{equ: proposed_frame1}
    \end{aligned}
\end{equation}
Furthermore, to reduce the interference of the pilot carriers over the data symbols, an optional guard interval is left around each pilot carrier $x_r^p$ leading to 

\begin{figure}[t!]
    \centering
    \includegraphics[width=0.35\textwidth]{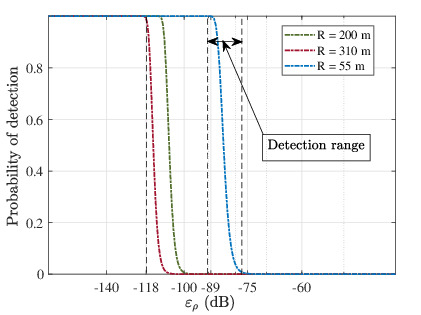}
    \caption{Probability of detection versus the windowing suppression coefficient $\varepsilon_\rho$ for different ranges, with $P_c = ¨P_r = 1W $.}
    \label{fig: PoD}
\end{figure}
\begin{figure}[t!]
    \centering
    \includegraphics[width=0.35\textwidth]{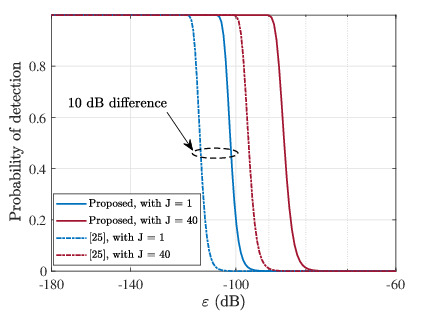}
    \caption{Probability of detection versus SIC residual coefficient $\varepsilon$  for $R=150$, with $P_c = ¨P_r = 1W $ and $\varepsilon_\rho = -80$ dB.}
    \label{fig: PoD_vs}
\end{figure}

\begin{equation}
    \begin{aligned}
    \mathbf{x}_r = \begin{cases}  
     x_r^p, & m=m_p \quad \quad p = 1, \ldots, N_r^p, \\ 
    0, & m \in \bigcup_{p=1}^{N_p}\left\{m: m=m_p+g, g \in \mathcal{G}_r\right\} \\
   x_r^d, & \textit{otherwise. }\end{cases}, 
      \label{equ: proposed_frame2}
    \end{aligned}
\end{equation}
where $\mathcal{G}_r$ represents the guard size around each pilot of the sensing signal. Consequently, the achieved spectral efficiency without channel coding is given by

\begin{equation}
    \begin{aligned}
    \eta_j=\frac{N_r^{eff}+N_c^{eff}}{N^{tot}} \log _2(1+\gamma_j^{\text{SINR}}),
        \label{equ: SE}
    \end{aligned}
\end{equation}
where $N_r^{eff}$ and $N_c^{eff}$ represent the data-carrying carriers of the radar and the communication signal, respectively, given by
\begin{subequations}
    \begin{align}
        N_r^{eff} &= N_r^{tot}-M_r \left[N_r^p(1+\mathcal{G}_r)+L_{cpp}\right], \label{eq: N_tot_a} \\
        N_c^{eff} &= N_c^{tot}-M_c \left[N_c^p+L_{cp}\right], \label{eq: N_tot_b}
    \end{align}
\end{subequations}
with $N_c^p$ denoting the number of pilots for each \ac{OFDM} symbol. \textcolor{black}{Compared to a unified OFDM-only design where all sensing functions reuse the OFDM frame without reserving AFDM resources, the SE is given by \cite{arous2024high}
\begin{equation}
    \begin{aligned}
    \eta_j^{conv}=\frac{N_{c,OFDM}^{eff}}{N_c^{tot}} \log _2(1+\gamma_j^{\text{SINR}}),
    \end{aligned}
\end{equation}
 where $N_{c,OFDM}^{eff}$ denotes the effective data-carrying subcarriers of the OFDM signal, and $N_c^{tot} = N_{c,OFDM}^{eff} + M_c \left[N_c^p+L_{cp}\right]$. It is clear that $N_{c,OFDM}^{eff} > N_c^{eff}+ N_r^{eff}$ in the proposed OFDM–AFDM system. Consequently, the SE of our design is strictly lower, such as
 \begin{equation}
     \frac{\eta_j}{\eta_j^{conv}}<1.
 \end{equation} }  
\textcolor{black}{Additionally, the uplink communication, as indicated in Fig. \ref{fig: block_diagram} can be conducted as in conventional IBFD systems with appropriate power-domain SIC.}

\begin{figure}[t!]
    \centering
    \includegraphics[width=0.35\textwidth]{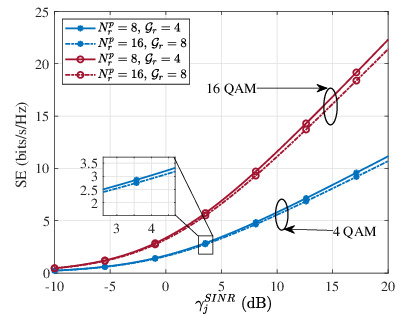}
    \caption{Spectral efficiency for different pilot and guard allocations, with $L_{cpp} = \frac{N_r}{8}$.}
    \label{fig: SE}
\end{figure}

\begin{figure}[t!]
    \centering
    \includegraphics[width=0.4\textwidth]{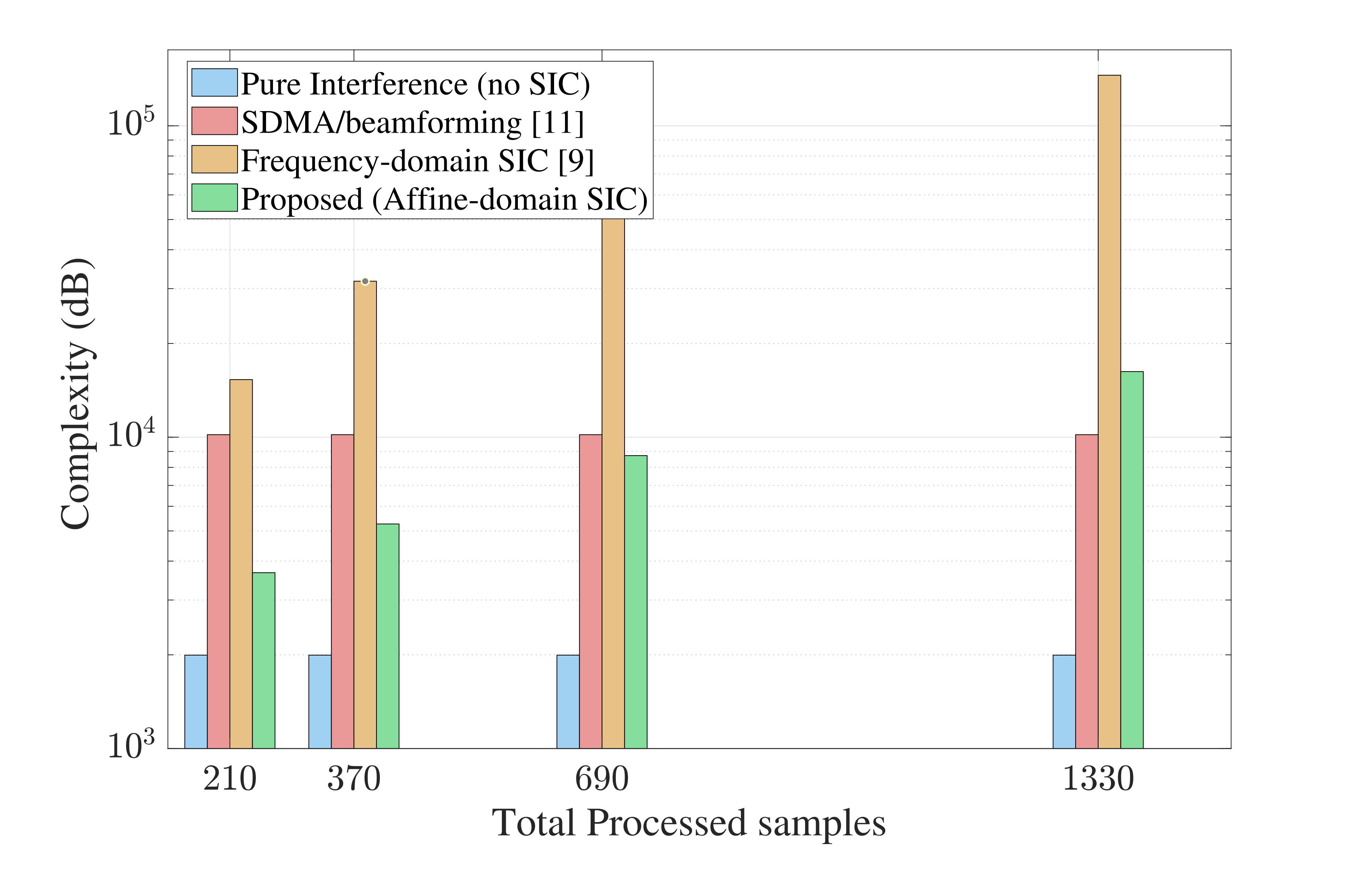}
    \caption{\textcolor{black}{System's computational complexity as function of the the total processed samples.}}
    \label{fig: comp}
\end{figure}

\begin{figure*}[t]
    \centering
    \subfigure[]{\includegraphics[width=0.23\textwidth]{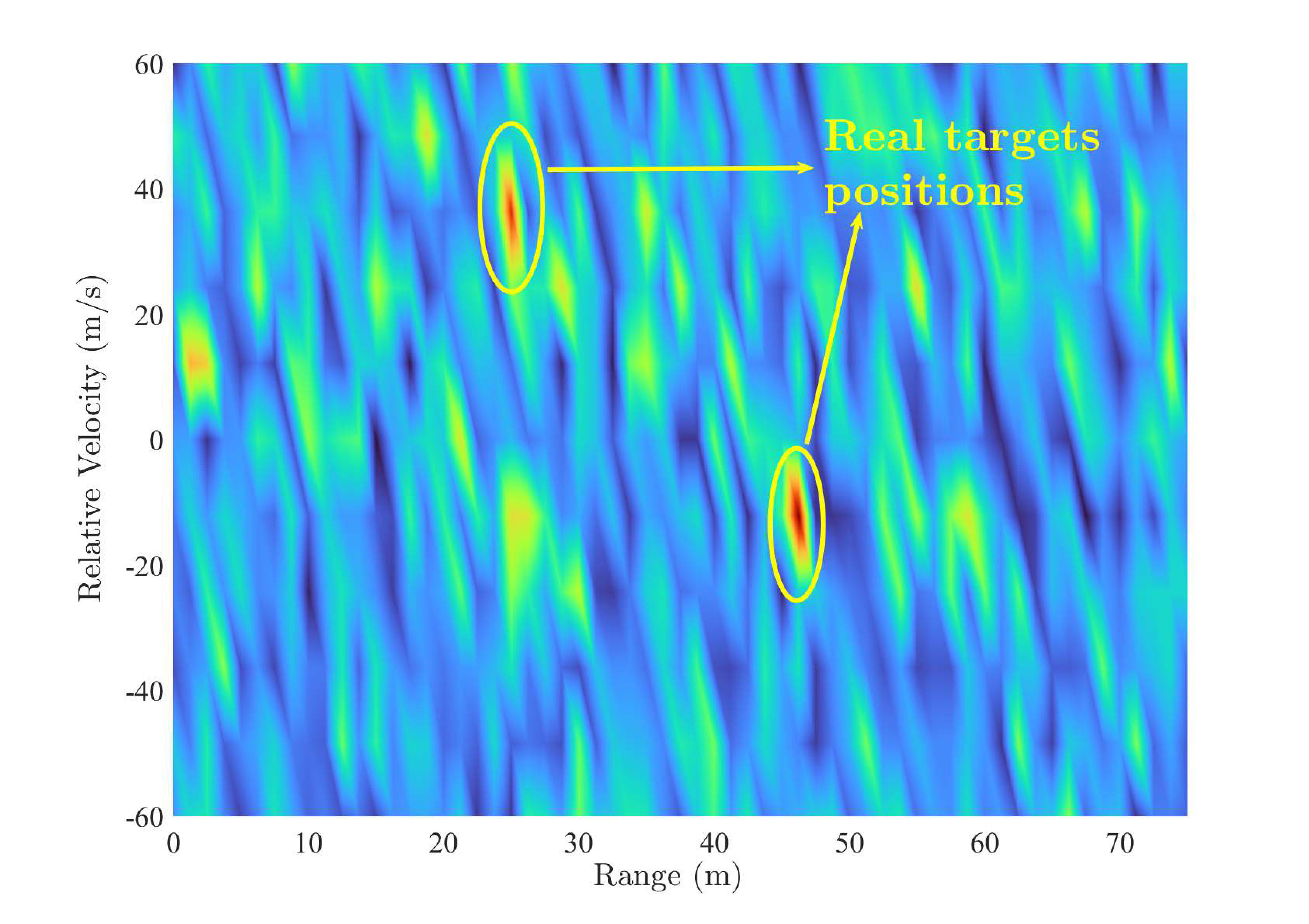}} 
   \hfill
    \subfigure[]{\includegraphics[width=0.23\textwidth]{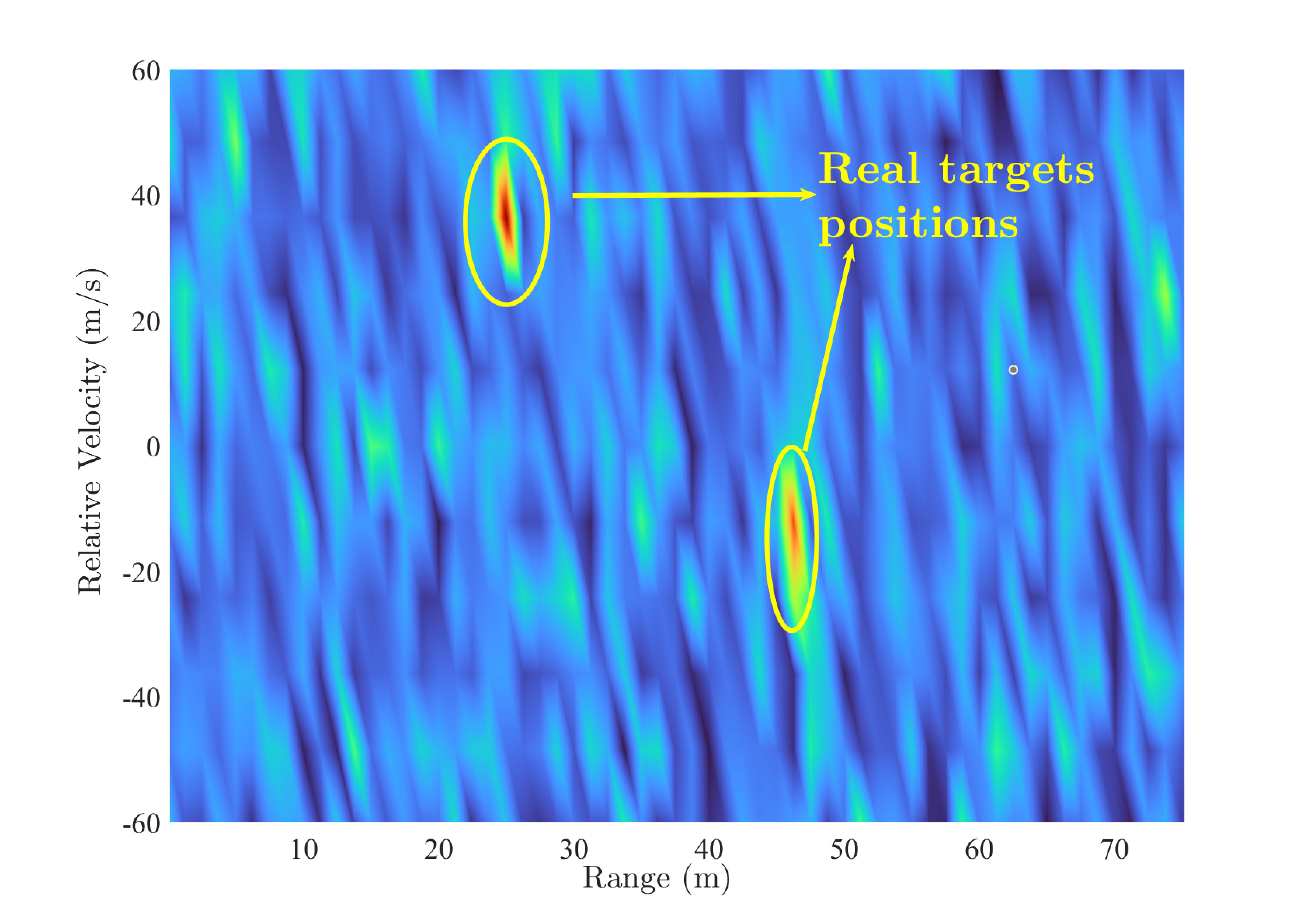}} 
   \hfill
       \subfigure[]{\includegraphics[width=0.23\textwidth]{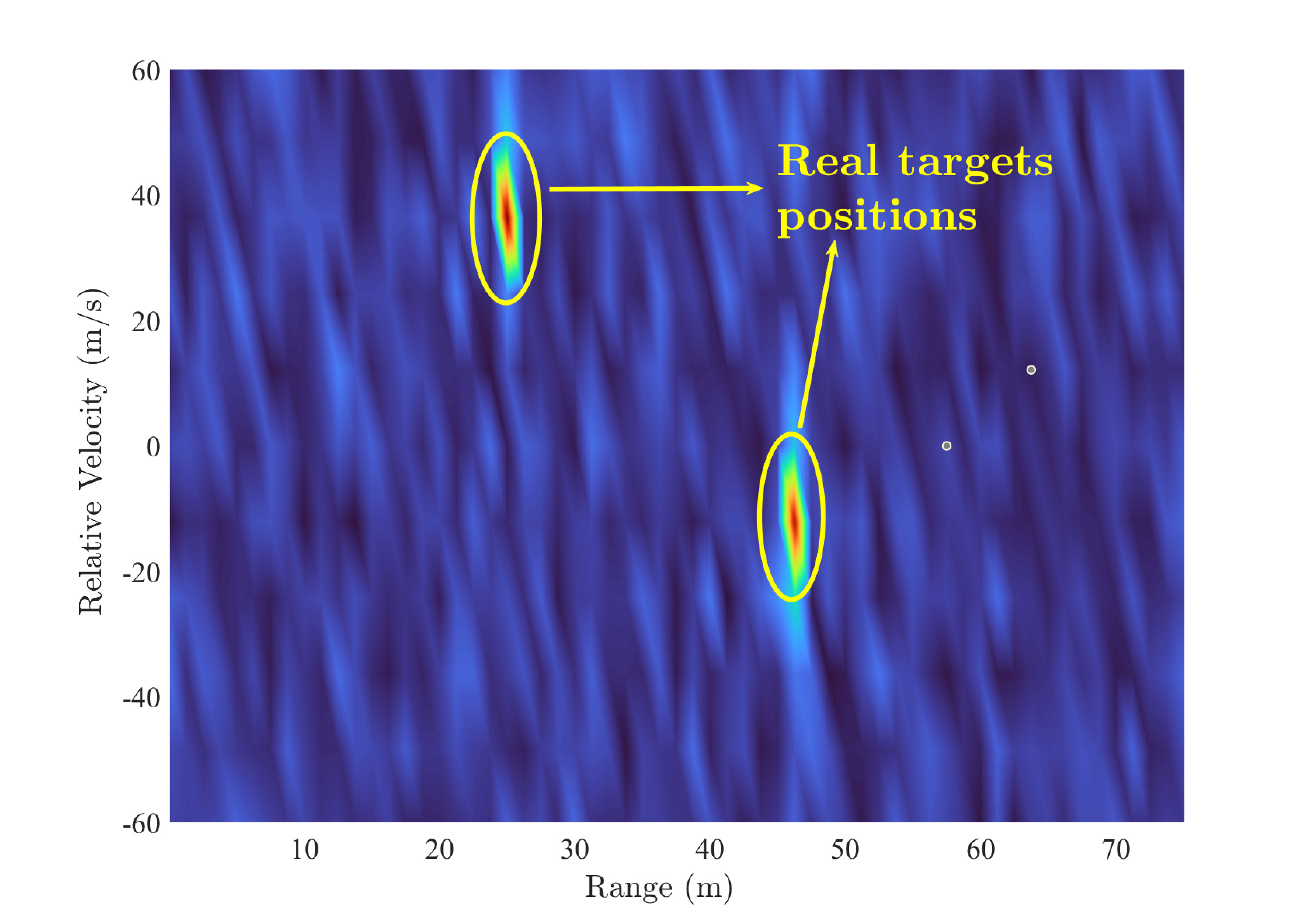}} 
 \hfill
    \subfigure[]{\includegraphics[width=0.23\textwidth]{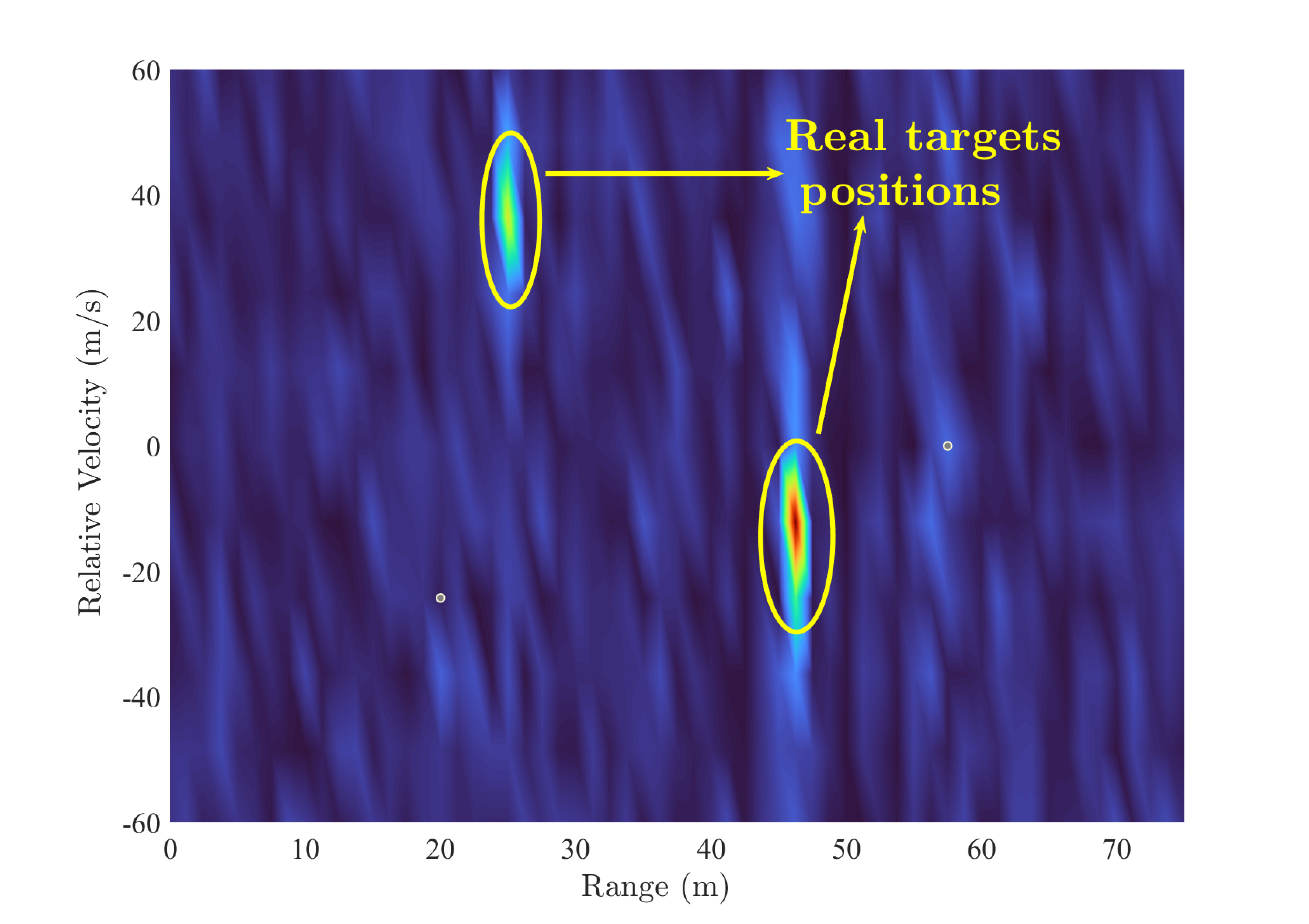}}
     \hfill

    \caption{RDM for single pilot frame with $N_r^p=1, \quad |x_r^p|^2=10$ dB: (a) Raw signal without SIC and windowing, (b) Only SIC in affine domain, (c) SIC with windowing ($\varepsilon_{\rho}=- 70$ dB) , and (d) SIC with windowing ($\varepsilon_{\rho}=-90$ dB).}
    \label{fig: RDM_one_pilot}
\end{figure*}

\begin{figure*}[t]
    \centering
    \subfigure[]{\includegraphics[width=0.23\textwidth]{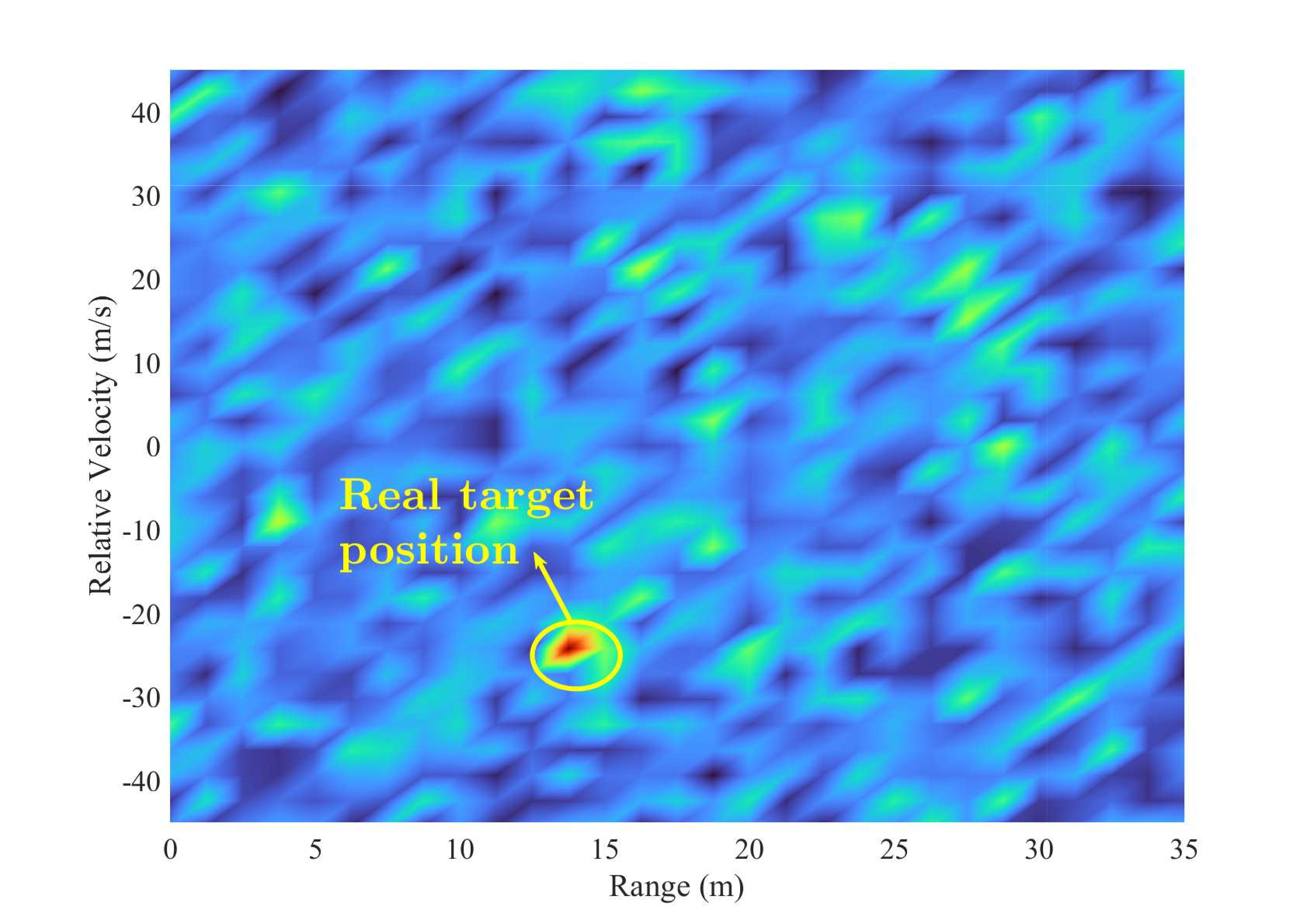}} 
    \hfill
    \subfigure[]{\includegraphics[width=0.23\textwidth]{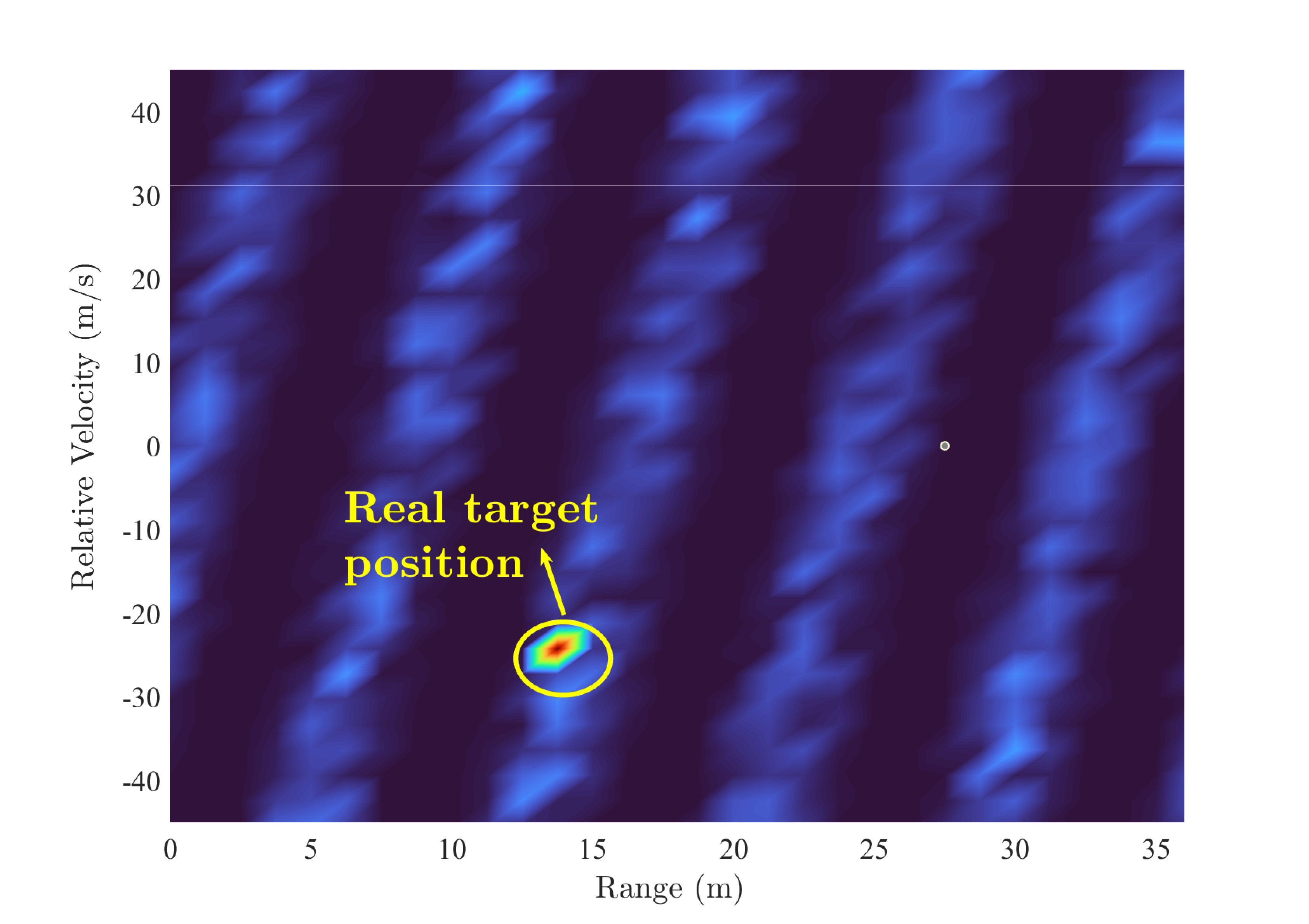}} 
   \hfill
    \subfigure[]{\includegraphics[width=0.23\textwidth]{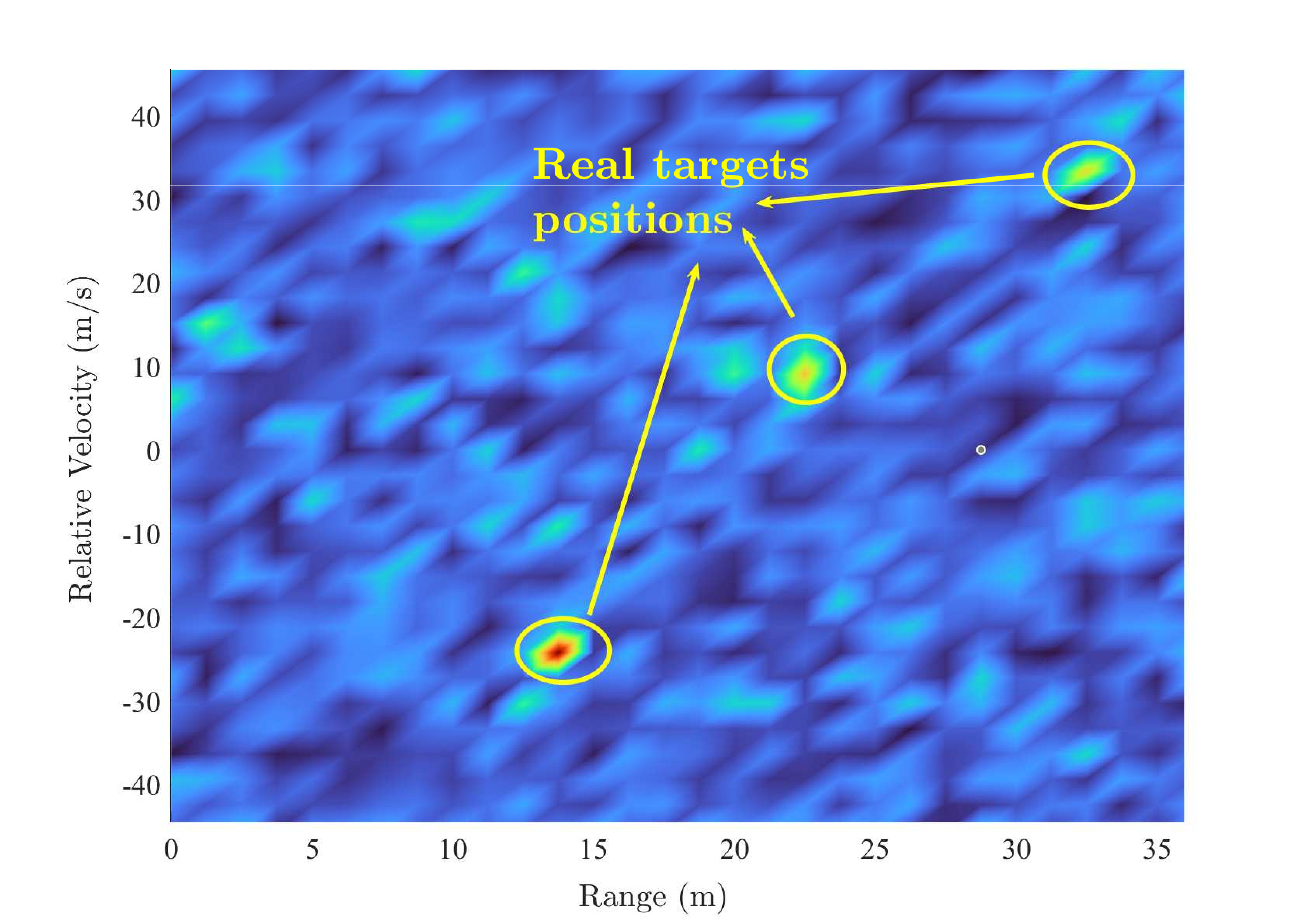}}
     \hfill
    \subfigure[]{\includegraphics[width=0.23\textwidth]{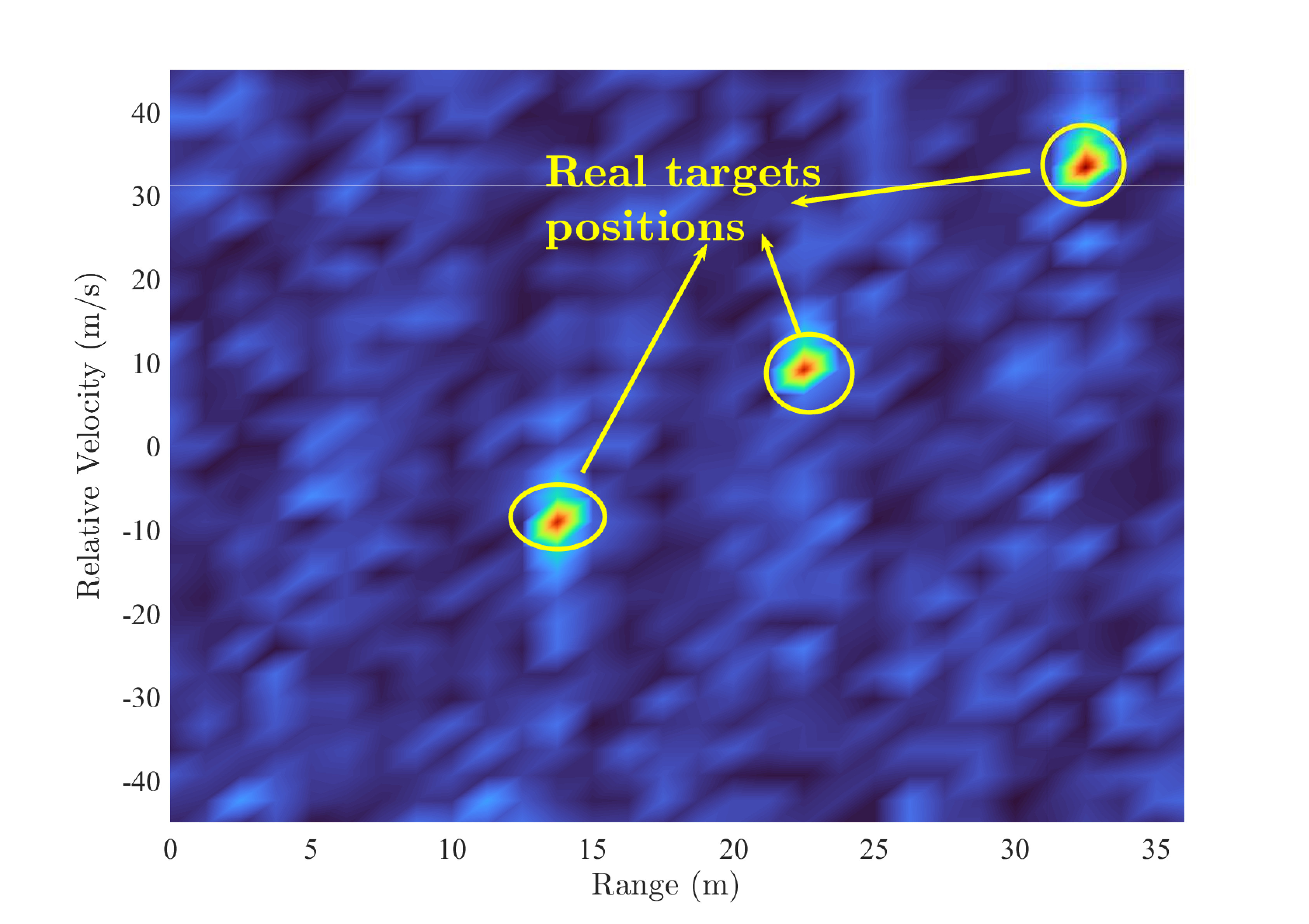}} 
 \hfill

    \caption{RDM for multi-pilot frame with $N_r^p=16, \quad |x_r^p|^2=2$ dB: (a) Single target with SIC only, (b) Single target with SIC and windowing ($\varepsilon_{\rho}=-90$ dB), (c) Multi-targets with SIC only, and (d) Multi-targets with SIC and windowing ($\varepsilon_{\rho}=-70$ dB).}
    \label{fig: RDM_multi_pilot}
\end{figure*}

\section{Simulation Results}
In this section, the performance of the proposed waveform domain IBFD-ISAC SIC is evaluated through numerical simulations, with the simulation parameters summarized in Table I. \textcolor{black}{For sensing, an \ac{EVA} channel model is adopted, with a dominant line of sight (LoS) having a complex amplitude of the radar return} modeled as $\alpha = \frac{G_t G_r \lambda^2 \sigma }{(4\pi)^3R^4}$.
\par In Fig. \ref{fig: PoD}, the probability of detection as a function of the windowing suppression $\varepsilon_{\rho}$ is plotted for different target's ranges. It can be seen that the iterative windowing provides an additional degree of freedom for enhancing the detection probability by selecting different suppression levels (where $\varepsilon_{\rho}$ depends on the implemented window). Due to the initial SIC in the affine domain, a moderate value of $\varepsilon_{\rho}$ around $-118$ dB is sufficient to detect distant targets at $R=310$, as noise suppression enables target distinction by adapting the detection threshold for weak targets. Moreover, close targets can be detected with reduced windowing levels of $\varepsilon_{\rho} \in [-89,75] \text{ dB}$. Additionally, a moderate $\varepsilon_{\rho}$ not only ensures a high detection probability but also reduces the complexity of windowing implementation, which makes the proposed system feasible under limited  system complexity. Alternatively, the probability of detection as a function of the SIC residual coefficient $\epsilon$ is illustrated in Fig. \ref{fig: PoD_vs} for $R = 150 \text{ m}$. For the same processed \ac{PRI}s, the proposed system can provide a high probability of detection compared to \cite{xiao2022waveform}, with a $10 \text{ dB}$ SIC residual difference. This is mainly due to the additional layer of windowing suppression within each \ac{PRI}, as indicated in \eqref{equ: SINR}.  
\par The proposed frame structure, as depicted in Fig. \ref{fig: SE}, demonstrates enhanced spectral efficiency, even when a substantial number of radar pilots are deployed. This is attributed to the fact that the communication functionality remains independent of the radar signal design. While a single radar pilot, i.e., $N^r_p = 1$, is theoretically sufficient for radar information extraction, the adopted frame configuration utilizes a greater number of pilots and incorporates appropriately designed guard intervals, to ensure effective separation between data and pilot channel responses. Nevertheless, the increase in the number of radar pilots has a negligible impact on spectral efficiency. Alternatively, the overall system complexity is mostly determined by \( N_s \), which is a function of the windowed samples. Since the iterative windowing process initially locks on the target bins, the number of required samples remains constrained. Then, even when \( N_{tot} \) is increased whether by incorporating additional radar symbols to enhance sensing performance or by introducing more communication symbols to improve spectral efficiency, the resulting increase in computational complexity remains minimal. \textcolor{black}{As shown in Fig. \ref{fig: comp}, the per-PRI complexity of the proposed affine-domain SIC grows only moderately as \(N^{tot}\) reflecting its FFT-type \(N\log N\) terms (dominant \(\mathcal{O}_1\) and \(\mathcal{O}_3\)) and a bounded, localized windowing stage \(\mathcal{O}_2\). Conversely, the frequency-domain SIC in [9] scales most steeply because it must search and estimate off-diagonal ICI taps across sidebands for every subcarrier, followed by reconstruction and subtraction over the full grid. For the SDMA technique, it adds an almost constant overhead since its spatial separability is largely independent of the baseband sample processing; however, its effectiveness presumes ideal conditions like accurate channel knowledge and feedback, spatial separability between the communication UE and sensing targets, and frequent beam sweeping to prevent beam overlap.}

\par Due to its superiority in providing pulse localization and interference suppression in chirp domain \cite{haif2024novel}, Kaiser window is adopted for the proposed ISAC system, defined as
\begin{equation}
w_{K W}[m]=\frac{J_0(0)\left(\bar{\beta} \sqrt{1-\left(\frac{m-\frac{N}{2}}{\frac{N}{2}}\right)^2}\right)}{J_0(\bar{\beta})},
\end{equation}

where $J_0(0)$ is the zeroth-order modified Bessel function of the first kind, and $\bar{\beta}$ is the suppression level of the window sidelobes. In Fig. \ref{fig: RDM_one_pilot}, the \ac{RDM} of the targets is illustrated for a single radar's pilot configuration. In this setup, the radar information is extracted by correlating the processed affine domain signal with the transmitted pilot to obtain the time domain channel. Subsequently, multiple AFDM radar symbols from the same \ac{PRI} are concatenated in the time domain, and a 2D-DFT is applied to generate the RDM. Fig. \ref{fig: RDM_one_pilot}(a) and (b) depict the RDM for the received signal before and after SIC in the affine domain, respectively. Fig. \ref{fig: RDM_one_pilot}(a) shows the \ac{RDM} of the raw signal, where due to the high levels of the SI power, high-intensity spikes emerge leading to ghost target detections even with an optimal detection threshold. This is mitigated by applying the proposed SIC in the affine domain as shown in Fig \ref{fig: RDM_one_pilot}(b). However, although OFDM spreading in the affine domain facilitates straightforward SIC,  a residual SI that amplifies the collective resulting noise persists. To mitigate that, an iterative windowing is applied over the SIC-processed signal using Kaiser window as shown in Fig. \ref{fig: RDM_one_pilot} (c) and (d), where it demonstrate the ability to suppress residual SI, even for lower value of $\varepsilon_\rho$ without increasing false alarm probability by creating new targets. 
\par To achieve higher resolution while processing fewer symbols per \ac{PRI} in the SI-free matrix \( \bar{\bold{R}} \), a multi-pilot frame structure is employed, as illustrated in Fig. \ref{fig: RDM_multi_pilot}. By leveraging the processing gain achieved through multiple pilots, the power allocated to each pilot is reduced from $10$ dB to $2$ dB, thereby significantly decreasing the signal's \ac{PAPR}. Unlike the single pilot frame, the multi-pilot approach applies a Kaiser window to all pilots channel responses in the affine domain following the SIC process, before transitioning to time domain processing. Since the windowing operation in the multi-pilot frame spans a larger number of samples in the affine domain, it effectively enhances noise suppression in the \ac{RDM}, as shown in Fig. \ref{fig: RDM_multi_pilot}(b). Furthermore, for multiple targets, the proposed system maintains robust target separation and detection capabilities due to the constructive behavior of pilots in the PCTD while requiring lower windowing suppression levels, as depicted in Fig. \ref{fig: RDM_multi_pilot}(d). 
\begin{figure}[t]
    \centering
    \subfigure[]{\includegraphics[width=0.35\textwidth]{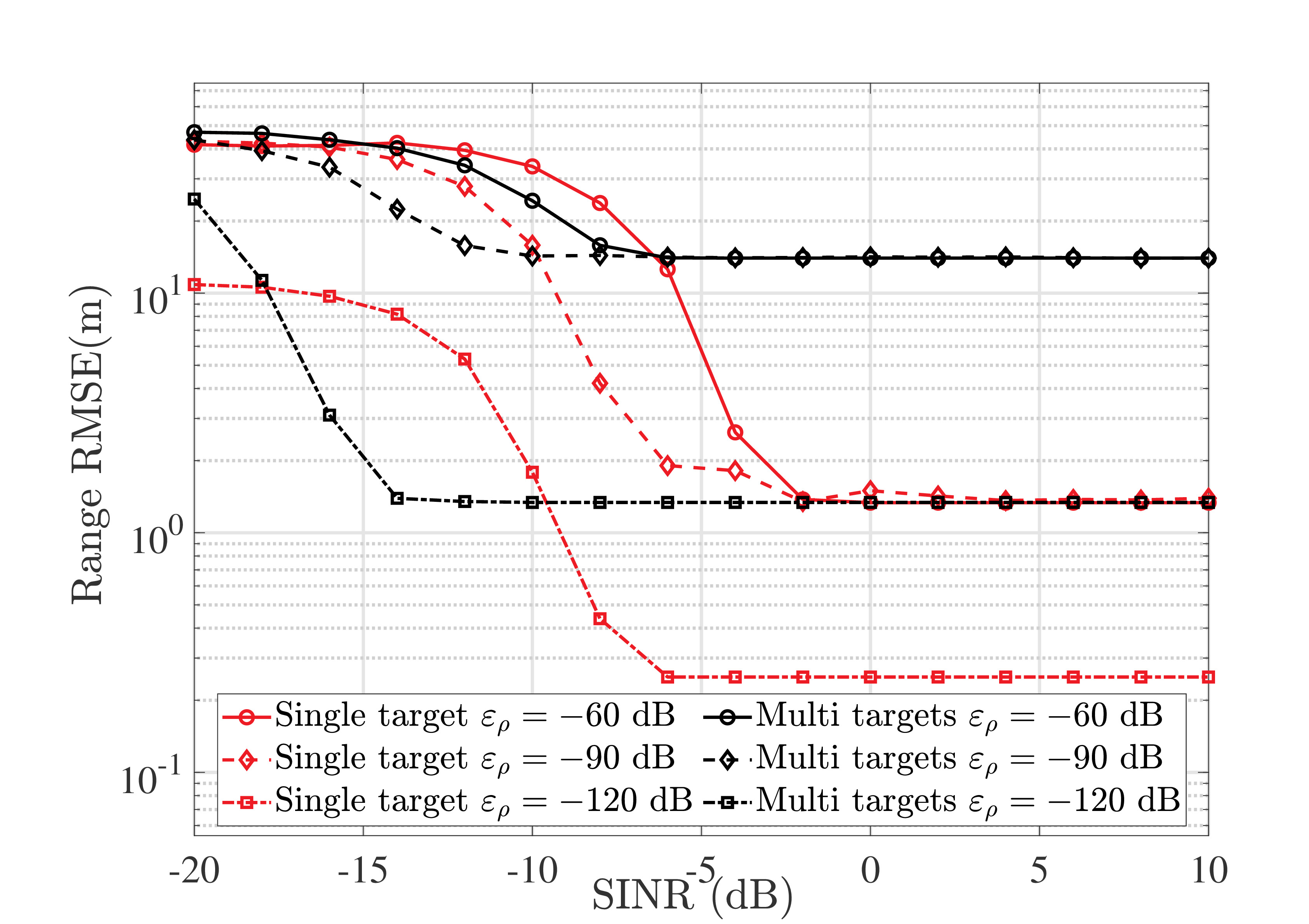}} 
    \hfill
    \subfigure[]{\includegraphics[width=0.35\textwidth]{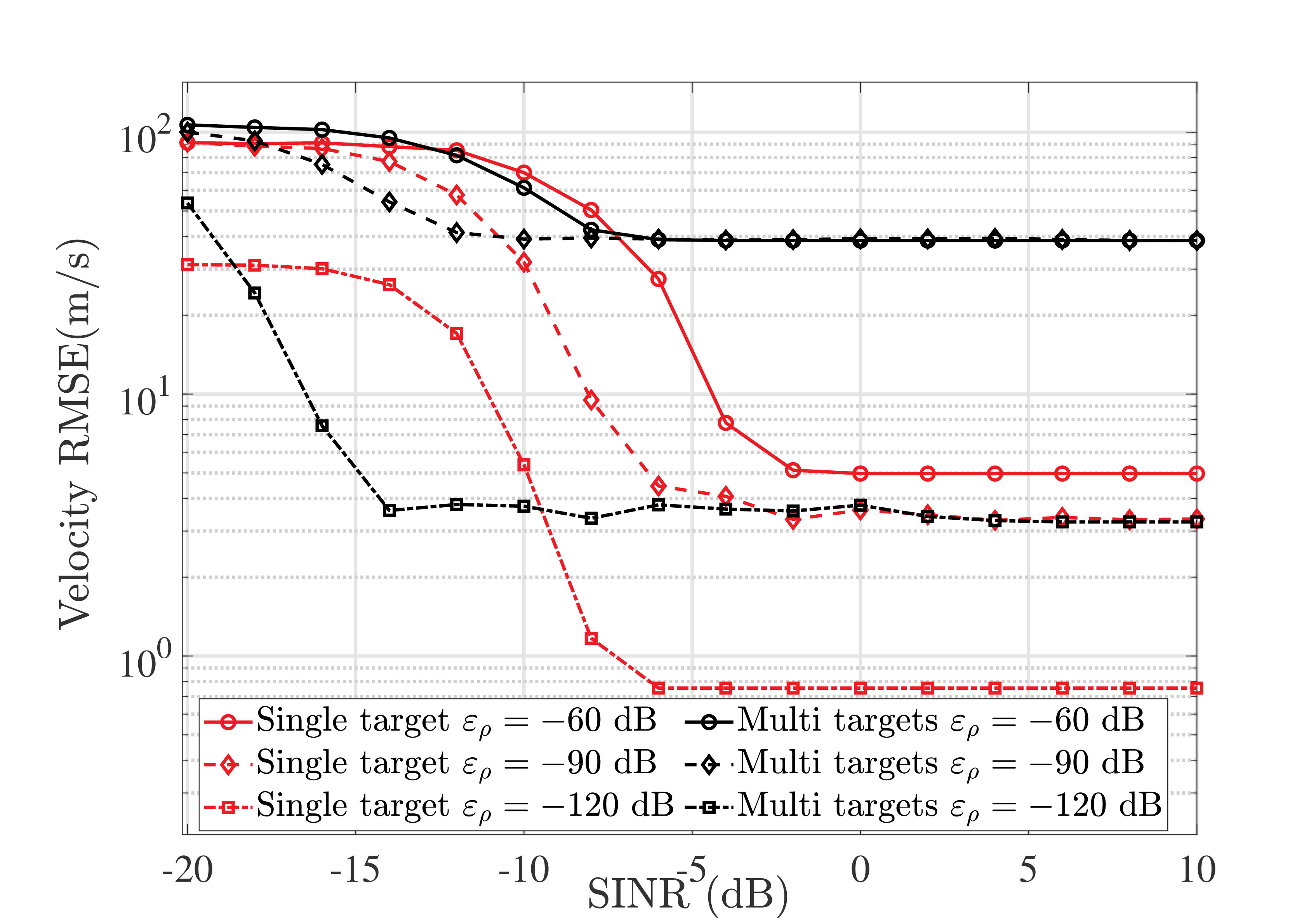}} 
   \hfill
    \caption{RMSE versus SINR for single pilot frame ($N_r^p=1, \quad |x_r^p|^2=10$): (a) Range and, (b) Velocity.}
    \label{fig: RMSE_one}
\end{figure}

\begin{figure}[t]
    \centering
    \subfigure[]{\includegraphics[width=0.35\textwidth]{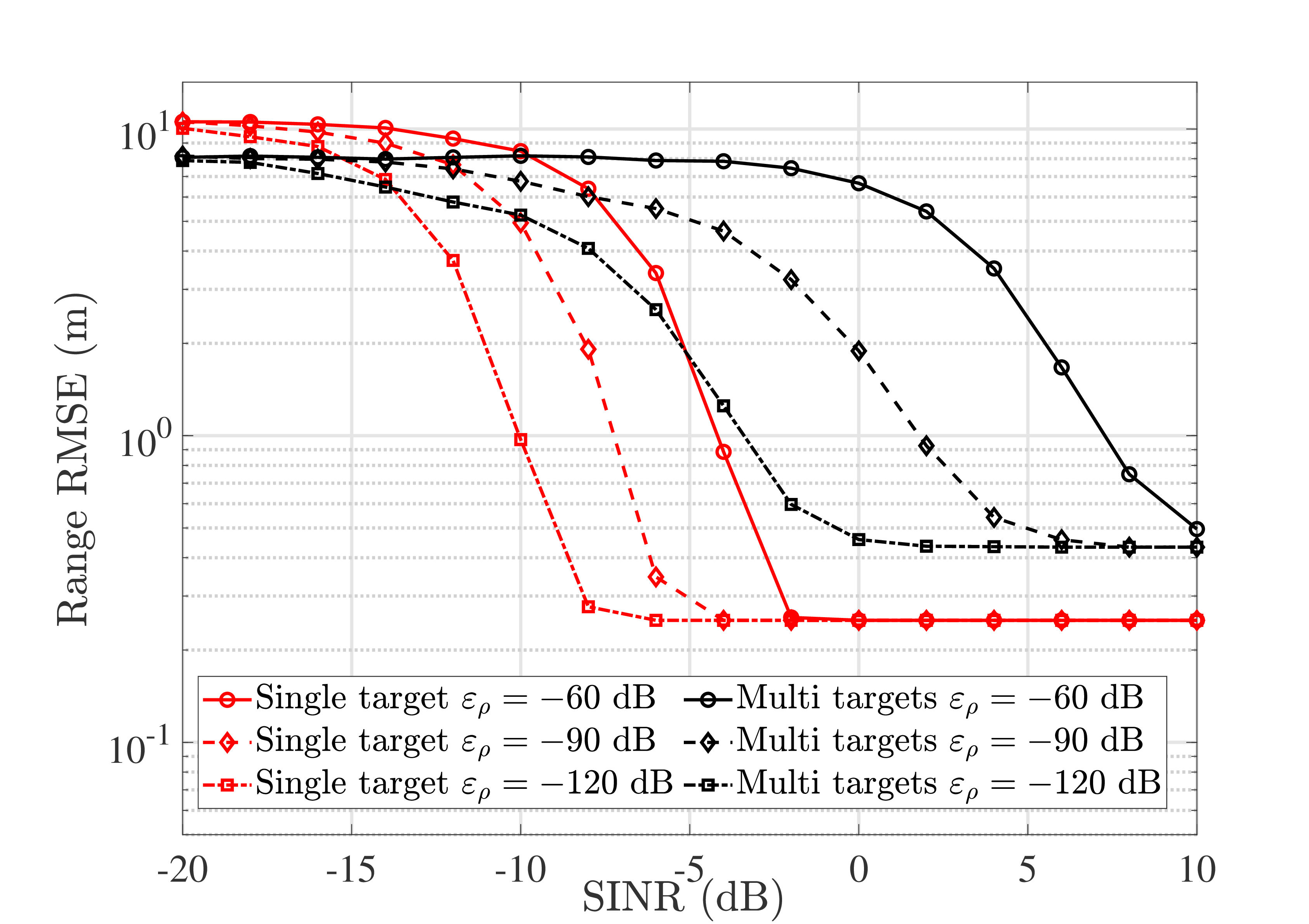}} 
    \hfill
    \subfigure[]{\includegraphics[width=0.35\textwidth]{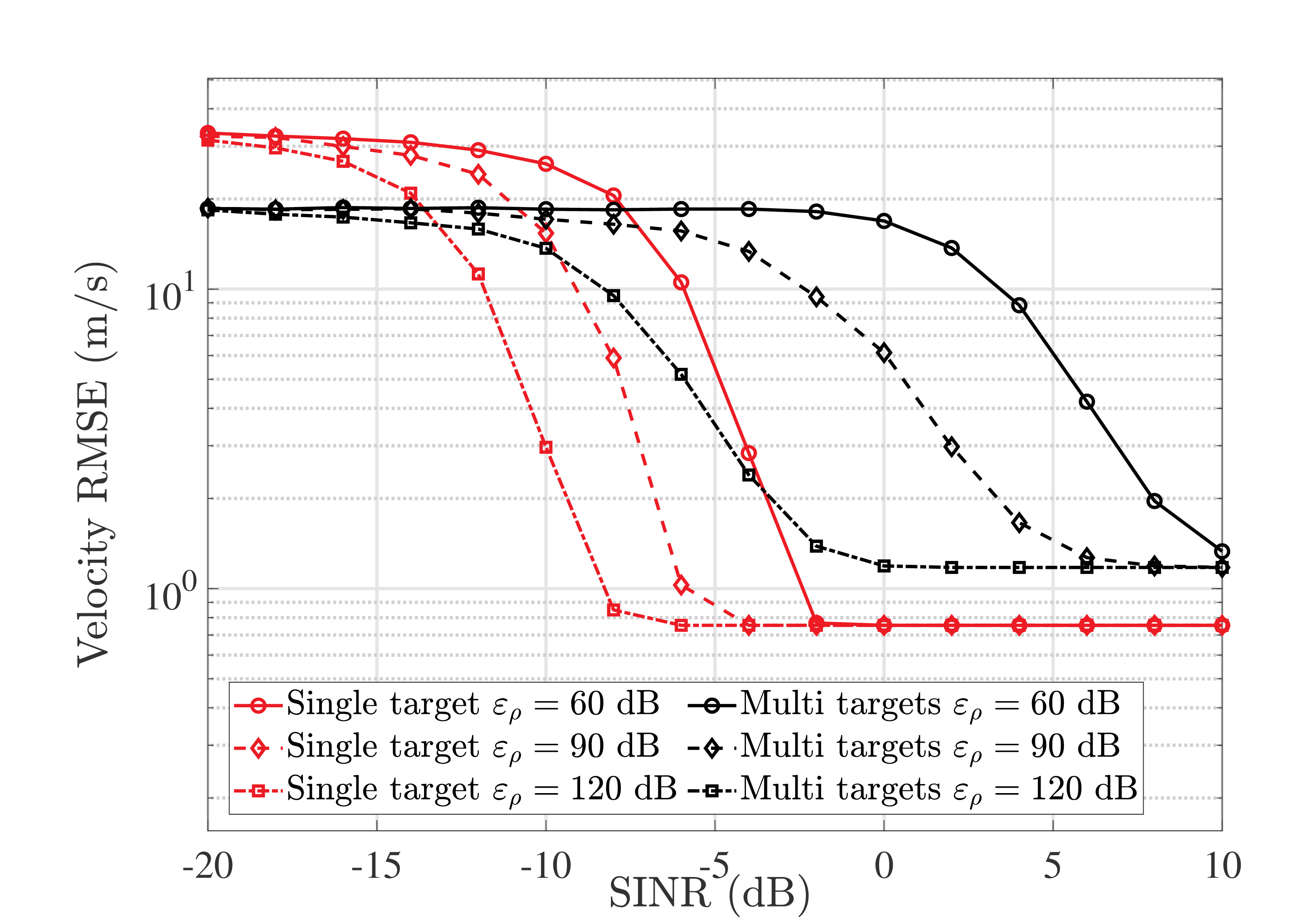}} 
   \hfill
    \caption{RMSE versus SINR for multi pilots frame ($N_r^p=16, \quad |x_r^p|^2=2, \quad z_p=2$): (a) Range and, (b) Velocity.}
    \label{fig: RMSE_multi}
\end{figure}

\par Fig. \ref{fig: RMSE_one} (a) and (b) depict the \ac{RMSE} of range and velocity, respectively, defined as  $RMSE(x) = \frac{1}{L} \sum^{L}_{i=1} \sqrt{\|x_i - \hat{x}_i\|^2}$ as a function of the \ac{SINR} under different windowing suppression levels. Specifically, in Fig. \ref{fig: RMSE_one} (a), the range RMSE for a single target at \(\varepsilon_\rho = -120\) dB converges at relatively low SINR values. However, as \(\varepsilon_\rho\) decreases, the range RMSE deteriorates due to the presence of residual SI. For multi-target scenarios, the single-pilot frame structure achieves lower RMSE values despite windowing, primarily due to the mutual interference among targets in the PCTD, where the energy of a single pilot disperses across the zero-padded matrix.  Similarly, in Fig. \ref{fig: RMSE_one} (b), the Doppler RMSE for multi-target scenarios follows a comparable pattern, with the localized Doppler peak in the PCTD remaining low even at high SINR and \(\varepsilon_\rho\) levels. This behavior arises because a single pilot lacks the constructive behavior in the PCTD, where the combined interference will have a strong power levels to at each entry of $\Bar{\bold{P}}_{j,zp}^{w_{\rho , l}} $
\begin{figure}[t]
    \centering
    \subfigure[]{\includegraphics[width=0.34\textwidth]{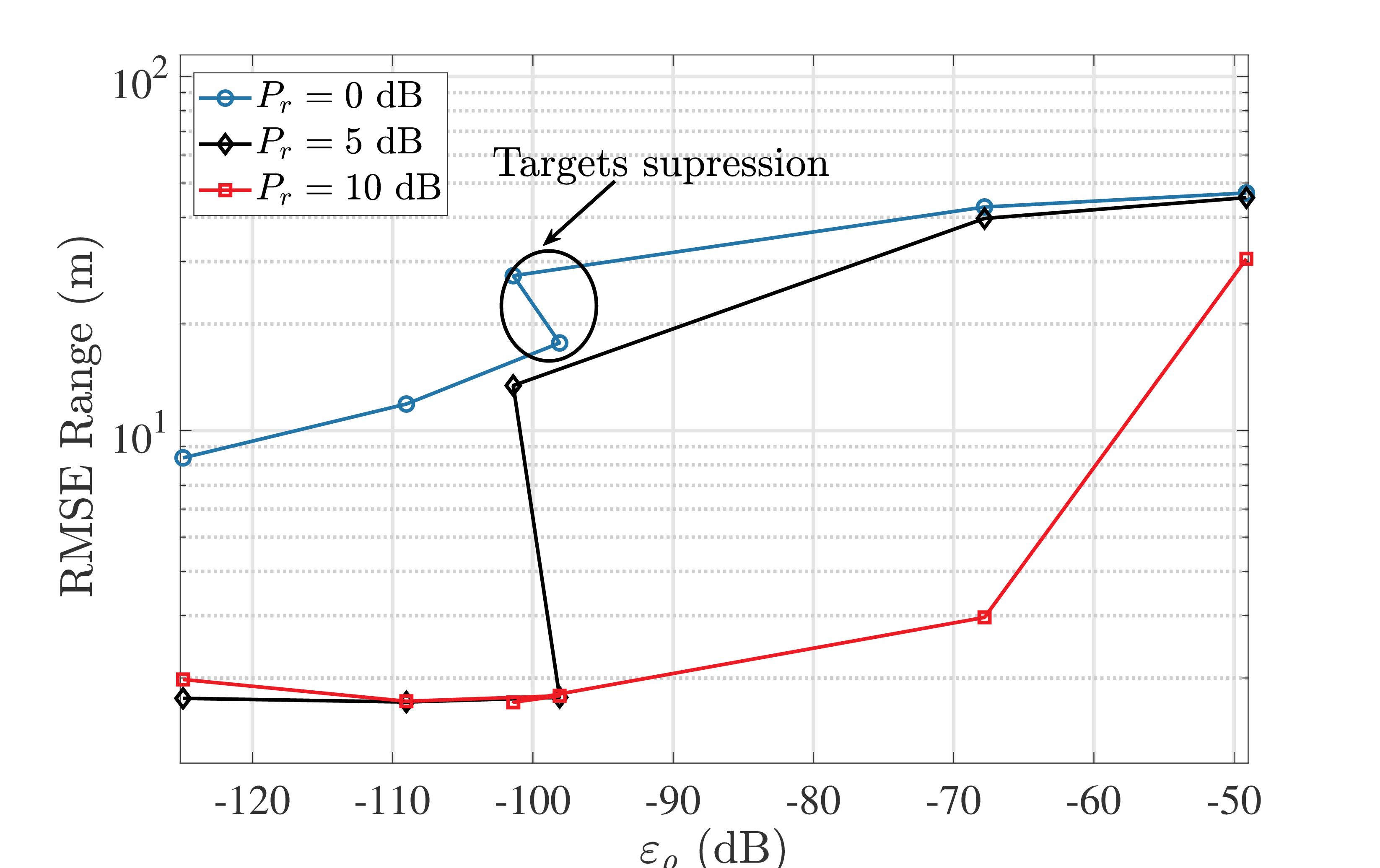}} 
    \hfill
    \subfigure[]{\includegraphics[width=0.34\textwidth]{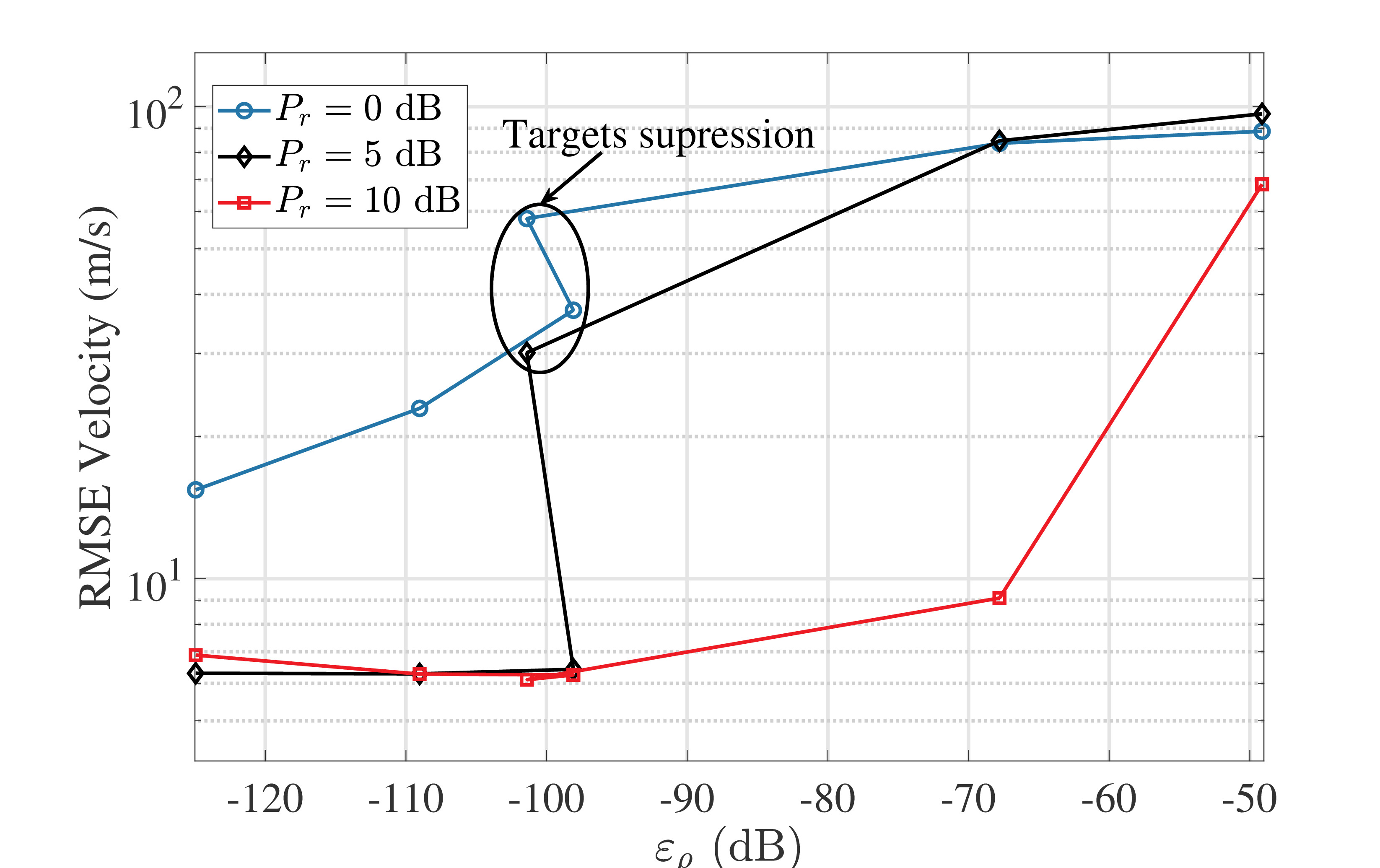}} 
   \hfill
    \caption{RMSE versus $\varepsilon_{\rho}$ under different radar signal power.}
    \label{fig: RMSE_window}
\end{figure}

\begin{figure}[t]
    \centering
    \subfigure[]{\includegraphics[width=0.34\textwidth]{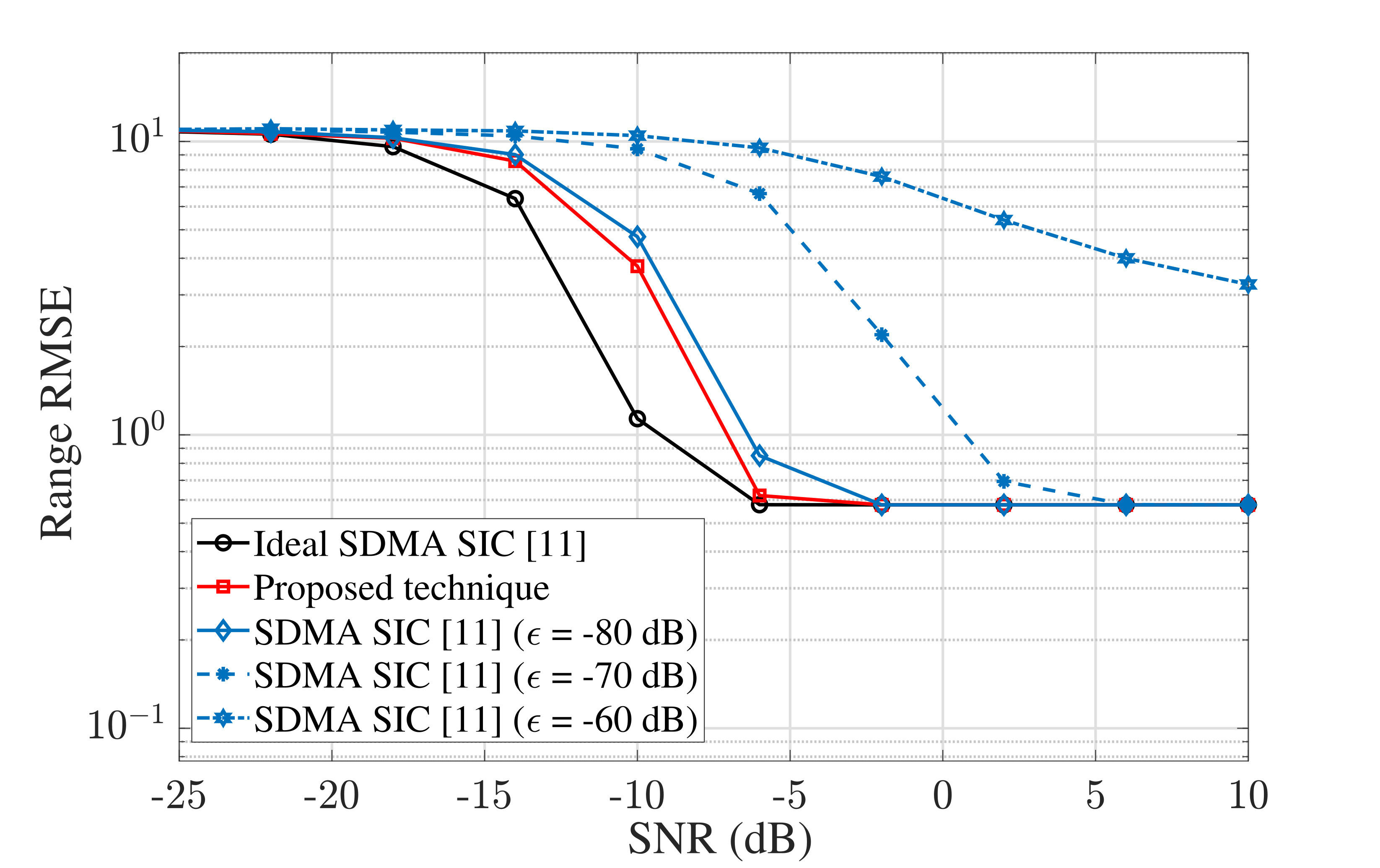}} 
    \hfill
    \subfigure[]{\includegraphics[width=0.34\textwidth]{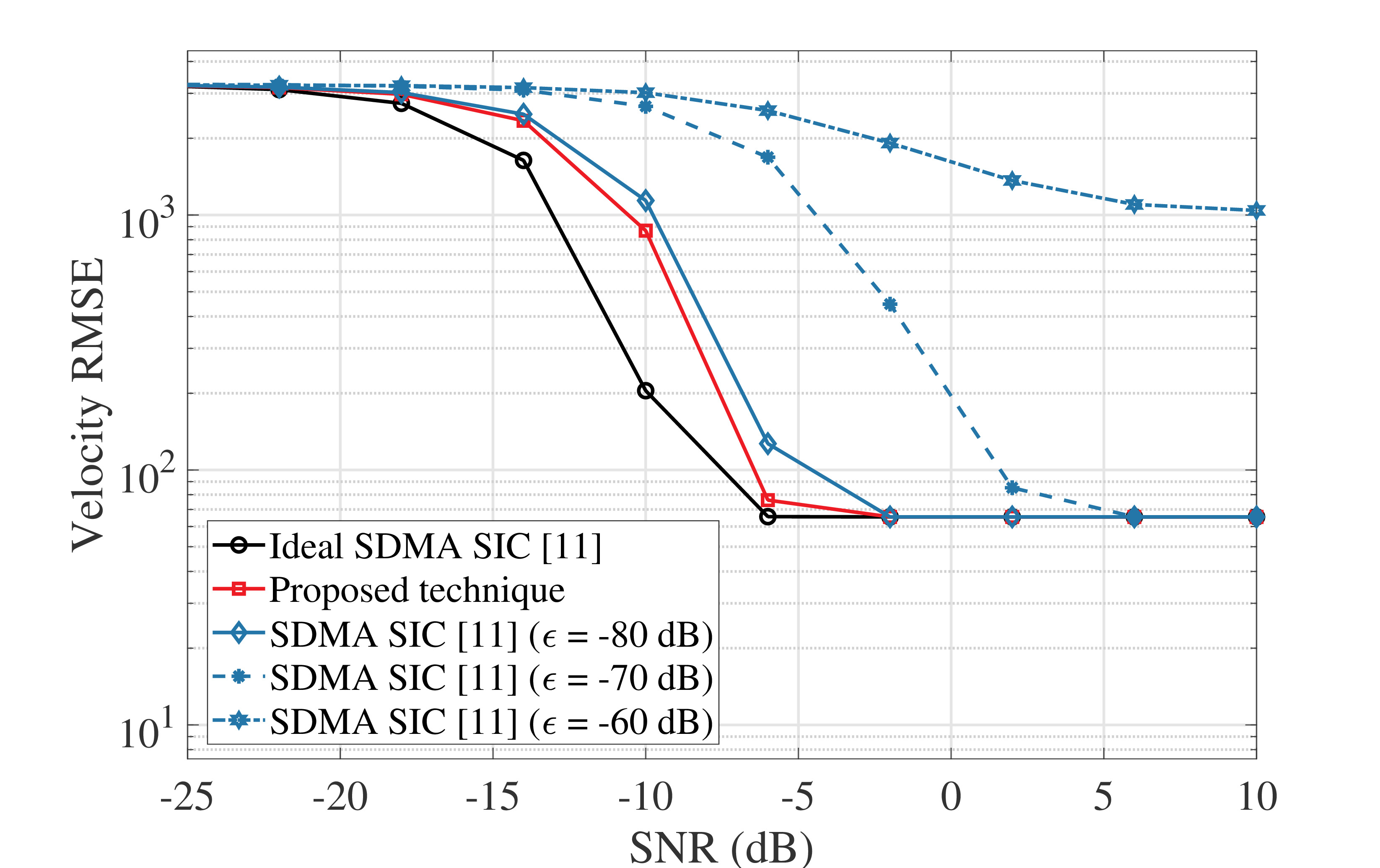}} 
   \hfill
    \caption{\textcolor{black}{range and velocity RMSE comparison with ideal and practical SDMA SIC.}
    \label{fig: RMSE_window_1}}
\end{figure}

For the multi-pilot frame, the accuracy of the estimated ranges and velocities depends on the variable $z_p$, which enables higher sensing resolution, as expressed below:
\begin{equation}
    \begin{aligned} 
        \hat{x}_{\text{multi}} = \text{argmax}(\Bar{\bold{P}}_{j,zp}^{w_{\rho , (\kappa,l)}}) \frac{1}{z_p\sqrt{N_s}}.
    \end{aligned}
\end{equation}
Fig. \ref{fig: RMSE_multi} (a) and (b) illustrate the achievable RMSE for multi-pilot structure for range and velocity, respectively. By leveraging the phase constructive properties of the chirp-carriers in the \ac{PCTD}, even moderate values of $\varepsilon_{\rho}$ achieve RMSE-bound values at low SINR for single and multi targets scenarios. This arises from the fact that the residual SI samples, after applying SIC in the affine domain combine destructively in PCTD, allowing the localized targets bins to show higher power levels in $\Bar{\bold{P}}_{j,zp}^{w_{\rho , l}} $ and  $\Bar{\bold{P}}_{j,zp}^{w_{\rho , \kappa}} $. Moreover, The RMSE for both range and velocity versus $\varepsilon_{\rho}$ is plotted in \ref{fig: RMSE_window}. The RMSE for both range and velocity converges to lower values as $\varepsilon_{\rho}$ increases, which is an expected outcome. However, once $\varepsilon_{\rho}$ exceeds $-100$ dB, no further improvement in RMSE is observed for high radar signal power. This indicates that the proposed technique serves as an opportunistic approach, enabling optimal sensing information by balancing radar signal power and window complexity. Notably, within the range of $\varepsilon_{\rho} \in [-90,-100]$ dB, certain spikes appear in the RMSE curves, primarily due to the suppression of nearby targets. That is because wider window sizes exhibit a degraded performance, while narrow windows precisely localize each target. Consequently, specific ranges suppress nearby targets, leading to irregular RMSE behavior.
\begin{figure}[t]
    \centering
    \subfigure[]{\includegraphics[width=0.34\textwidth]{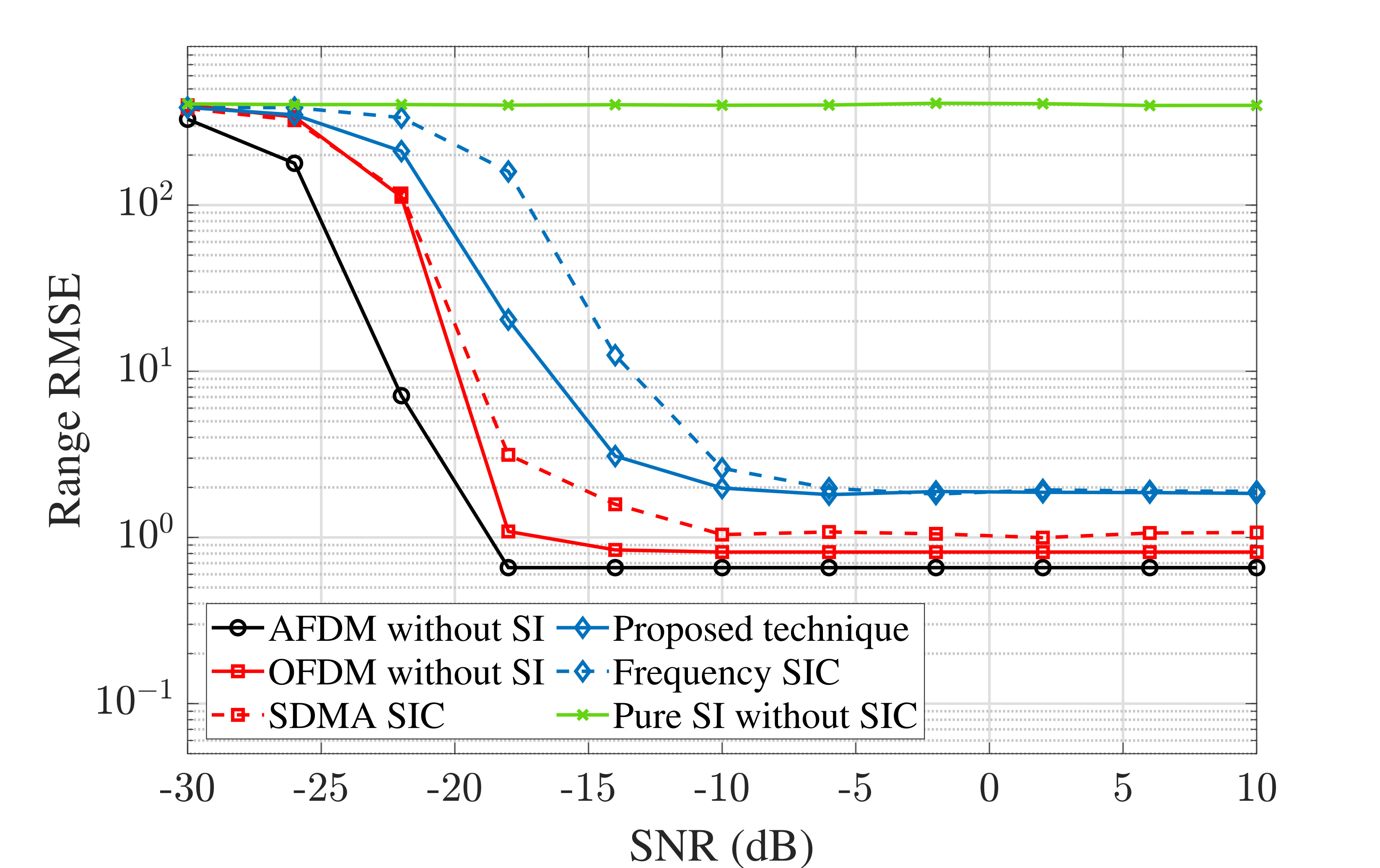}} 
    \hfill
    \subfigure[]{\includegraphics[width=0.34\textwidth]{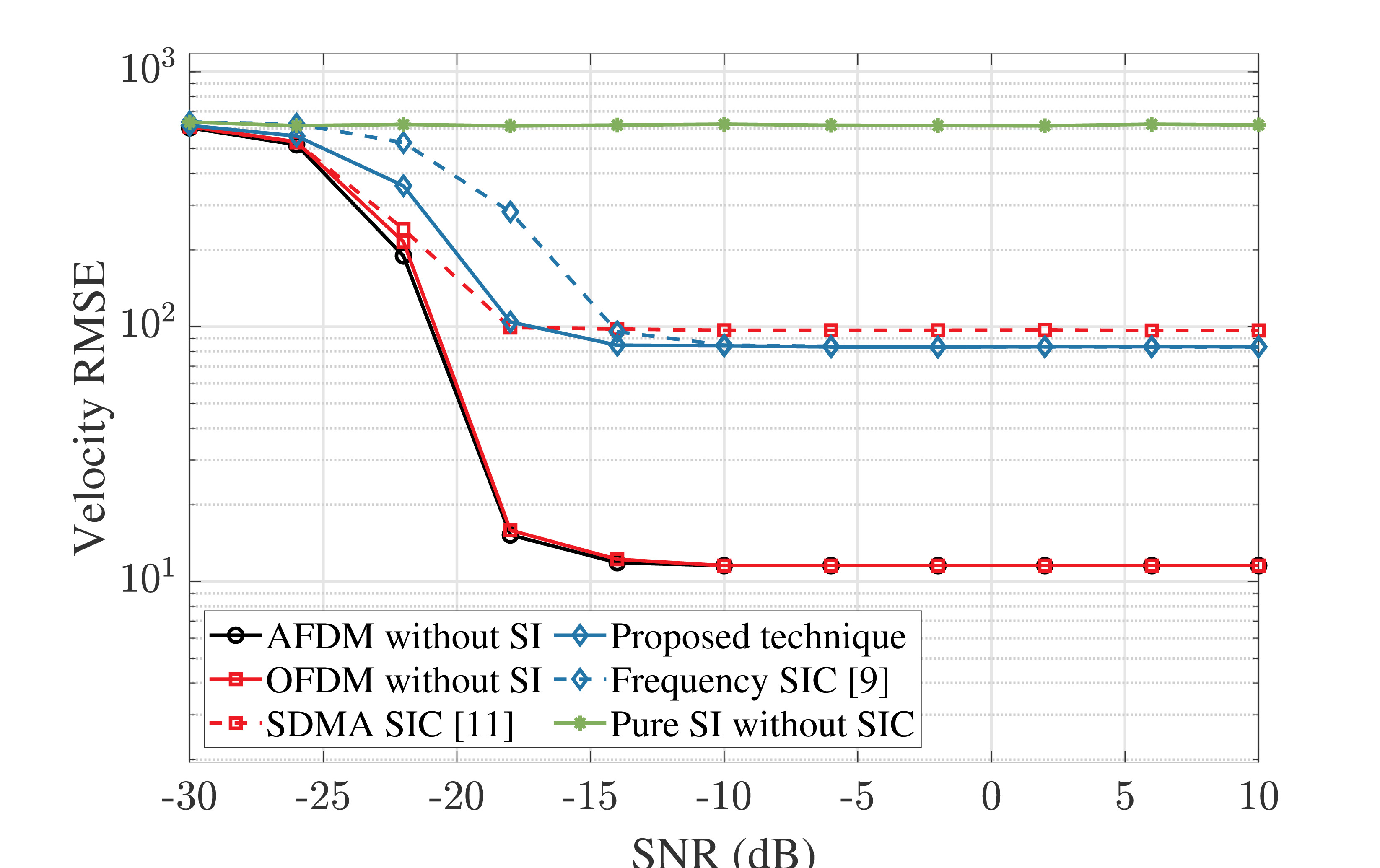}} 
   \hfill
    \caption{\textcolor{black}{Range and velocity RMSE for different SIC techniques.}}
    \label{fig: RMSE_window_2}
\end{figure}
 \textcolor{black}{Fig. 12  demonstrates the performance of the proposed technique with multiple pilots frame $|x^p_r|^2 = 5$ dB and $\epsilon_p = 90$ dB, in comparison with the SDMA technique [11] at different $\epsilon$ levels for multi target sensing. The proposed affine-domain SIC matches ideal SDMA through the transition region from approximately $-15$ to $-5$ dB and attains the same low floor, whereas practical SDMA degrades under realistic CSI/beam/hardware constraints, showing a higher error floor with slow convergence at higher SNR values. Hence, the proposed approach achieves near-ideal range/velocity accuracy without stringent spatial assumptions. In addition, Fig. 13  depicts the performance of the proposed technique with multiple pilots frame $|x^p_r|^2 = 2$ and $\epsilon_p = 90$ dB, versus baselines for multi-target sensing in range and velocity RMSE. The waveform-domain SIC converges to the SI-free OFDM's performance more rapidly than frequency-domain SIC, which retains residual SI due to accumulated channel-estimation errors during iterative ICI cancellation. Simultaneously, it shows a near $0$ dB difference compared to the ideal SDMA SIC technique.}
\section{Conclusion}
This work introduces a novel approach for SIC in IBFD-ISAC systems by leveraging waveform-domain projections, to provide an additional degree of freedom for SIC. Specifically, we propose an adaptive ISAC frame generation mechanism that alternates between OFDM for communication and AFDM for radar operations, utilizing the generalized multicarrier waveform generation framework. We analytically demonstrate that the OFDM waveform, which induces strong SI on the radar echo, exhibits a statistical behavior similar to AWGN in the affine domain, thereby facilitating its cancellation. To further mitigate residual SI, an adaptive iterative low-complexity windowing scheme is employed. Then, an additional time-domain SI spreading phase is introduced to minimize the impact of delay and Doppler variations. The proposed framework reduces the need to process large PRIs over the CPI matrix to a single PRI, by leveraging the shift-like behavior of the diminishing Doppler phase in the PCTD. Simulation results validate the effectiveness of the proposed scheme in detecting multiple targets under strong SI and low SINR levels, which is further enhanced by adaptively optimizing the windowing coefficients. By leveraging this adaptability, a high probability of detection is achieved even in the presence of significant residual SI. 
\bibliographystyle{IEEEtran}
\bibliography{Mendeley.bib}
\end{document}